%Paper: hep-th/9503212
%From: Eva Silverstein <silver@puhep1.Princeton.EDU>
%Date: Thu, 30 Mar 1995 03:53:58 GMT
%Date (revised): Sun, 30 Apr 1995 11:57:00 -0400

% Upper-case    A B C D E F G H I J K L M N O P Q R S T U V W X Y Z
% Lower-case    a b c d e f g h i j k l m n o p q r s t u v w x y z
% Digits        0 1 2 3 4 5 6 7 8 9
% Exclamation   !           Double quote "          Hash (number) #
% Dollar        $           Percent      %          Ampersand     &
% Acute accent  '           Left paren   (          Right paren   )
% Asterisk      *           Plus         +          Comma         ,
% Minus         -           Point        .          Solidus       /
% Colon         :           Semicolon    ;          Less than     <
% Equals        =           Greater than >          Question mark ?
% At            @           Left bracket [          Backslash     \
% Right bracket ]           Circumflex   ^          Underscore    _
% Grave accent  `           Left brace   {          Vertical bar  |
% Right brace   }           Tilde        ~

\input harvmac.tex
\overfullrule=0pt
\def\bt{\bf{\overline{27}}}
\def\bar{\overline}
\def\bq{{\overline Q_{+}}}

\def\s{\sigma}
\def\as{|\sigma|}
\def\at{|t_R|}
\lref\distgreene{J. Distler and B. Greene,``Some Exact Results on the
Superpotential from Calabi-Yau Compactifications'', {it Nucl. Phys.}
{\bf B309} (1988) 295.}
\lref\grass{E. Witten, ``The Verlinde Algebra and the Cohomology
of the Grassmannian'', IASSNS-HEP-93/41, hepth/9312104.}
\lref\rigid{P. Aspinwall and B. Greene, ``On the Geometrical Interpretation
of N=2 Superconformal Theories'', CLNS-94-1299, hepth/9409110.}
\lref\coleman{S. Coleman, ``More on the Massive Schwinger Model'',
{\it Ann. Phys.} {\bf 101} (1976) 239.}
\lref\stress{E. Silverstein and E. Witten, ``Global U(1) R-Symmetry
and Conformal Invariance of (0,2) Models'', {\it Phys. Lett.} {\bf } }
\lref\jdnotes{J. Distler, ``Notes on (0,2) Sigma Models'',}
\lref\wessbagg{J. Wess and J. Bagger, {\it Supersymmetry and Supergravity},
2nd ed., Princeton University Press, 1992.}
\lref\banksetal{T. Banks, L. Dixon, D. Friedan, and
E. Martinec, ``Phenomenology and Conformal Field Theory, or
Can string theory predict
the weak mixing angle?'', {\it Nucl. Phys.} {\bf B299} (1988) 613.}
\lref\dkfirst{J. Distler and S. Kachru, ``(0,2) Landau-Ginzburg Theory'',
{\it Nucl. Phys.} {\bf B413} (1994) 213, hepth/9309110.}
\lref\spectrum{S. Kachru and E. Witten, ``Computing the Complete Massless
Spectrum of a Landau-Ginzburg Orbifold'', {\it Nucl. Phys.}{\bf B407}
(1993) 637, hepth/ 9307038.}
\lref\dksecond{J. Distler and S. Kachru, ``Singlet Couplings and (0,2)
Models '', {\it Nucl. Phys.} {\bf B430} (1994) 13, hepth/9406090.}
\lref\daddaetal{A. D'Adda, A.C. Davis, P. Di Vecchia, and
P. Salomonson, ``An Effective Action for the Supersymmetric
$CP^{n-1}$ Model'', {\it Nucl. Phys.} {\bf B222} (1983) 45.}
\lref\cdgp{P. Candelas, X. de la Ossa, P. Green, and L. Parkes,
``A pair of Calabi-Yau Manifolds as an Exactly Soluble Conformal
Field Theory'', {\it Nucl. Phys.} {\bf B359} (1991) 21.}
\lref\egyang{T. Eguchi and S.K. Yang, ``N=2 Superconformal Models
as Topological Field Theories'', {\it Mod. Phys. Lett.} {\bf A5}
(1990) 1693.}
\lref\dixon{L. Dixon, ``Some World-Sheet Properties of Superstring
Compactifications, in Orbifolds and Otherwise'', Lectures given at
the 1987 ICTP Summer Workshop in High Energy Physics and Cosmology.}
\lref\ww{X.G. Wen and E. Witten, ``World Sheet Instantons and the
Peccei-Quinn Symmetry'', {\it Phys. Lett.} {\bf 166B} (1986) 397.}
\lref\dsww{M. Dine, N. Seiberg, X.G. Wen, and E. Witten, ``Nonperturbative
Effects on the String Worldsheet'' I and II, {\it Nucl. Phys.}
{\bf B278} (1986) 769 and {\bf B289} (1987) 319.}
\lref\distgreene{J. Distler and B. Greene, ``Aspects of (2,0) String
Compactifications'', {\it Nucl. Phys.} {\bf B304} (1988) 1.}
\lref\miracles{M. Dine and N. Seiberg, ``Are (0,2) Models String Miracles?'',
 {\it Nucl. Phys.} {\bf B306} (1988) 137.}
\lref\jd{J. Distler, ``Resurrecting (0,2) Models'',
{\it Phys. Lett.} {\bf B188} (1987) 431.}
\lref\phases{E. Witten, ``Phases of N=2 Theories in Two Dimensions'',
{\it Nucl. Phys.} {\bf B403} (1993) 159.}
\lref\bagwit{E. Witten and J. Bagger, ``Quantization of Newton's
Constant in Certain Supergravity Theories'', {\it Phys. Lett.}
{\bf 115B} (1982) 202.}
\lref\gsw{M. Green, J. Shwarz, and E. Witten, {\it Superstring Theory,
vol. 2}}
\lref\bcov{M. Bershadsky, S. Cecotti, H. Ooguri, and C. Vafa,
``Kodaira-Spencer Theory of Gravity and Exact Quantum String Amplitudes'',
{\it Commun. Math. Phys.} {\bf 165} (1994) 311.}
\lref\mirgep{E. Silverstein, ``Miracle at the Gepner Point'',
PUPT-1519, hepth/9503150.}
\lref\candelas{P. Candelas, ``Yukawa Couplings of (2,1) Forms'',
{\it Nucl. Phys.} {\bf B298} (1988) 458.}
\lref\toplg{C. Vafa, ``Topological Landau-Ginzburg Models'',
{\it Mod. Phys. Lett.} {\bf A6} (1991) 337.}
\lref\topsig{E. Witten, ``Topological Sigma Models'', {\it Comm. Math
Phys.} {\bf 118} (1988) 411.}
\lref\mirtop{E. Witten, ``Mirror Manifolds and Topological Field Theory'',
in {\it Essays on Mirror Manifolds}, ed. S-T Yau, International
Press (1992) 120.}
\lref\topgrav{E. Witten, ``On the Structure of the Topological Phase of
Two-Dimensional Gravity'', {\it Nucl. Phys.} {\bf B340} (1990) 281.}
\lref\agnt{ Antoniadis, Gava, Narain, and Taylor, ``Topological Amplitudes
in String Theory'', {\it Nucl. Phys.} {\bf B413} (1994) 162.}
\lref\gepnotes{D. Gepner, ``Lectures on N=2 String Theory'',
Lectures given at the Spring School on Superstrings, Trieste,
Italy, 3-11 April, 1989, PUPT-1121.}
\lref\ttstar{S. Cecotti and C. Vafa, ``Topological-
Anti-Topological Fusion'' {\it Nucl. Phys.} {\bf B367} (1991) 359.}
\lref\cvetic{M. Cvetic, ``Exact Construction of (0,2) Calabi-Yau
Manifolds'', {\it Phys. Rev. Lett.} {\bf 59} (1987) 2829.}
\Title{\vbox{\hbox{IASSNS-HEP-95/22, PUPT-1533}
\hbox{\tt hep-th/9503212}
}}
{\vbox{\centerline{CRITERIA FOR CONFORMAL INVARIANCE}
\centerline{OF $(0,2)$ MODELS}}}
\centerline{Eva Silverstein\footnote{$^\dagger$}
{silver@puhep1.princeton.edu}}
\smallskip\centerline{Joseph Henry Laboratories}
\centerline{Jadwin Hall}
\centerline{Princeton University}\centerline{Princeton, NJ 08544 USA}
\smallskip\centerline{and}
\smallskip\centerline{Edward Witten\footnote{$^*$}
{witten@sns.ias.edu}}
\smallskip\centerline{School of Natural Sciences}
\centerline{Institute for Advanced Study}
\centerline{Olden Lane}
\centerline{Princeton, N.J. 08540}
\vskip .1 in
It is argued that many linear (0,2) models
flow in the infrared to conformally invariant
solutions of string theory.
The strategy in the argument is to show that the effective
space-time superpotential must vanish because there is no place
where it can have a pole.
This conclusion comes from either of two different analyses,
in which the Kahler class or the complex structure of the gauge
bundle is varied, while keeping everything else fixed.
In the former case, we recover from the linear sigma model the usual simple
pole in the ${\bf \bar {27}}^3$ Yukawa coupling but show that an analogous
pole does not arise in the couplings of gauge singlet modes.

\Date{03/95} %replace this line by \draft  for preliminary versions
	     %or specify \draftmode at some point
%\draft

\newsec{Introduction}

%--recall Dixon proof for flatness of (2,2) moduli and why it fails here

%--recall worldsheet instantons

%--review of relevant low-energy SUGRA which shows spacetime superpotential P
%is section of holomorphic line bundle with positive curvature

%--review of (0,2) LSM

$N=1$ spacetime supersymmetry in string theories requires
at least (0,2) supersymmetry on the worldsheet \banksetal.
The bulk of effort in studying string vacua has, however, gone into
(2,2) theories, in part because
they are easier to study; in particular, sufficient conditions for
conformal invariance are better understood.
For instance, Dixon
\dixon~ showed using the superconformal Ward identities that (2,2)-preserving
gauge singlet fields are moduli of string theory, a result which
holds to all orders in string loop perturbation theory.
The proof used the left-moving supersymmetries crucially, and indeed
it was noted early on \ww\dsww~ that unlike (2,2) models, (0,2) sigma models
are susceptible to worldsheet instanton corrections to the vacuum
energy.  Certain (0,2) sigma models
have been argued to avoid destabilization by instantons
\jd\cvetic\miracles\distgreene, but the conditions
required for these arguments
have been very special.

Linear sigma models give a new way to study the
parameter spaces of (0,2) and (2,2) theories \phases.  A very
large class of models can be represented by linear sigma models;
this includes, for instance, arbitrary complete intersections in
weighted projective spaces (or more general toric varieties).
Distler and Kachru have used linear sigma models
to study interesting classes
of (0,2) vacua in their Landau-Ginzburg
phases \dkfirst~ and showed, for example, that
a wide class of $E_6$-invariant $(0,2)$ Landau-Ginzburg models
were conformally invariant \dksecond.

In this paper we give new methods to argue for conformal invariance
of those (0,2) models that can be represented by linear sigma
models.  We focus for illustration on the $E_6$-invariant $(0,2)$
deformations of the quintic, but we believe that the considerations
are far more general.

The basic idea is to study the space-time superpotential
as a function of the moduli on which it depends; insofar as the moduli
space is compact and the superpotential is not identically zero,
the superpotential must have poles somewhere.\foot{A
holomorphic function without poles on a compact complex manifold
would have to be constant.  The superpotential is not really a holomorphic
function but a section of a line bundle of negative curvature \bagwit,
so it cannot even be constant: if there are no poles, it must vanish.}
Poles can only arise
when the parameters are taken to values at which the compactness of
the target space is lost because some bose fields can go to infinity.
At large field strength, quantum corrections to the classical theory
are small and calculable, so the possible poles can be located.
Moreover, the polar parts of the various couplings can be explicitly
computed.  We will analyze the behavior as either the Kahler class
or the complex structure of the bundle is varied and show in each case
that (i) the linear sigma model gives a natural compact parameter space,
(ii) the places where the sigma model breaks down can be concretely
described, and (iii) the relevant couplings do not have poles at those
places.

The paper is organized as follows.  In section 2, we explain heuristically
why linear sigma models may be special and why, in fact, conformal
invariance should possibly be understood on more elementary grounds
than we will explore here.
In section 3, we analyze the singularity locations of the linear
sigma model. In section 4, we perform computations near the singularity
of the Kahler moduli space,
recovering the usual pole in Yukawa couplings
of charged fields and showing that no pole arises for the $E_6$ singlet
modes.  Together with the compactness of the Kahler moduli
space of the linear sigma model,
this gives our first argument for conformal invariance.
In section 5, we analyze the behavior as a function of the complex
structure of the manifold and bundle, showing again the usual pole
for the charged fields but the absence of the pole for the
singlets.  The argument here can  be  carried out as in section 4.
In section 6 we briefly
discuss more general models.

\newsec{More Elementary Considerations}

In trying to compute non-zero space-time superpotentials for moduli fields
in $(0,2)$ linear sigma models, we encountered certain difficulties
which make us suspect that the problem of conformal invariance in
these models should perhaps be understood in a more elementary way
than we will eventually present.  We will here sketch some of these issues.

To compute the spacetime superpotential from a worldsheet theory, one
needs to know what interactions are determined by a given superpotential.
The general structure of Yukawa couplings coming from a superpotential
$W$ is as follows (using the conventions
of \wessbagg):
\eqn\stlag{L=\dots -e{\rm~ }\exp(K/2)\biggl\{W\bar\psi_a\bar\s^{ab}\bar\psi_b
+{i\over 2}\sqrt{2}{\rm D}_iW\chi^i\s^a\bar\psi_a
+{1\over 2}{\sl D}_i{\rm D}_jW\chi^i\chi^j + h.c.\biggr\}}
Here the $\psi_a$ is the gravitino, the $\chi^i$ are fermions
from chiral multiplets (their scalar partners will be called $A^i$),
and $K$ is the Kahler potential.
One way to study the superpotential is through its
covariant derivatives using for example the last term in \stlag; this
is what we will do in the bulk of this paper.  But a simpler option
seems to present itself here:  we could compute $W$ directly by
evaluating the $<\bar\psi_a\bar\s^{ab}\bar\psi_b A^i>$ coupling, which would
yield $-e ({1\over 2}{\partial K\over {\partial A^i}})
\exp(K/2) W$ evaluated at the vacuum expectation values
of the scalar fields $A_i$.
To compute this amplitude in string theory we compute the correlation
function of the corresponding vertex operators in the
two-dimensional quantum field theory.

This correlation function can be restricted by considering the right-moving
$U(1)$ $R$ charge.
As will be discussed more fully in section 3, the transformation
properties of a mode under spacetime gauge and Lorentz symmetries determines
its $U(1)$ charges.  The gravitino generates spectral flow, and
(in the canonical ghost picture)
has internal right $U(1)$ charge $q_R^{gravitino}=3/2$.
Spacetime scalar fields $A^i$ have $q_R^{scalar}=\pm 1$
(aside from the dilaton, which comes from the gravitational
sector of the theory and has zero internal right U(1) charge).
Therefore
the above amplitude does not conserve the internal right-moving $U(1)$
and the worldsheet correlation function vanishes trivially.

One might question whether the above argument is circular.  We have
used some of the usual properties of string solutions and vertex operators,
and perhaps the argument only proves that when one does have a string
solution, then the space-time superpotential is $W=0$.
In general, if one does not have a classical solution of string theory,
one does not know in what kind of world-sheet theory the computation
is to be performed.

This is where linear sigma models may be relevant.
Linear sigma models give definite quantum field theories that
flow from known (free) fixed points in the ultraviolet, and which
(under mild and familiar conditions) at least appear to flow
to conformal field theories in the infrared.
Moreover, many of their essential properties are known.
For instance,
the argument above mainly used the $R$ symmetry, which is valid even
away from criticality in appropriate linear models.

It might appear that if one does not have a conformal field theory,
one does not know what is meant by the gravitino and scalar vertex
operators in the above computation.  However, those vertex operators
are all chiral primaries (of the right-movers), which makes it possible
to identify them with states in the $\bar Q_+$ cohomology
(vertex operators of the half-twisted model) even away from
criticality.  Thus it appears that the above computation of $W=0$
makes sense without an {\it a priori} assumption of conformal invariance.

By contrast, consider a general Calabi-Yau manifold $X$ with a stable
holomorphic vector bundle $V$.  One can try to use this data to
determine a conformal field theory, and this is at least approximately
valid, modulo instanton corrections, near the field theory limit.
However, one does not know any {\it exact} quantum field theory,
conformally invariant or not, determined by this data.
In particular, one does
not know if these theories can be ``cut off'' in a way that preserves
the right-moving $R$ symmetry.  If in fact, because of instanton effects,
a general $(X,V)$ corresponds to a theory that is {\it not} conformally
invariant, and flows to {\it weak} coupling in the infrared, then
we may be in difficulty: it might be that any ultraviolet theory
that would flow to this renormalization group orbit is strongly coupled,
and we have no way to know if the $R$ symmetry can be maintained.

Thus, we believe that a possible picture is that a general $(X,V)$
does not correspond to any cut-off independent, $R$-invariant quantum
field theory in the ultraviolet or any conformal field theory in
the infrared.  On the other hand, those particular $(X,V)$
that can be realized as linear sigma models -- though vast in number,
they are a tiny fraction of abstract $(X,V)$'s\foot{The same assertion
is true for $(2,2)$ models: though a huge number of algebraic varieties come
from linear sigma models, generic ones do not.  Calabi-Yau manifolds
that do not come
from linear sigma models are perhaps less familiar as they
require more sophisticated algebraic
geometry for their construction and analysis.
See \ref\hubsch{T. Hubsch, {\it Calabi-Yau Manifolds: A Bestiary For
Physicists} (World-Scientific, 1992).}, pp. 103, 204 for a construction
of one such example.} --
do come from definite, known $R$-invariant theories
in which the above computation, giving $W=0$, can be performed.

Since we are in fact not sure whether the above line of reasoning
is valid, we will proceed in the rest of this paper to a more technical
discussion.  In passing, we will note also that we could study $(0,2)$ linear
sigma models from the point of view of analyzing non-perturbatively
the non-renormalization of the world-sheet
superpotential and twisted superpotential
(by methods that are familiar in four dimensions \ref\seiberg{N. Seiberg,
``The Power Of Holomorphy: Exact Results in 4-D SUSY Field Theories'',
RU-94-64 (1994) hepth/9408013.}).  Though we believe that such an analysis
would go through, we will not go down that road since the implications
of such an analysis for conformal invariance and the space-time superpotential
are not clear to us.  If such a relation could be understood, this
route might again give a more direct and elementary treatment of
conformal invariance of linear $(0,2)$ models than we will give
in this paper.

\newsec{Singularities}

%--singularities in the LSM:  unbounded bosonic zero-modes
%(emphasize this as the only source of divergences that we understand,
%given reasonable assumptions about metric and normalization...)

%--geometrical description of conifold singularity of the manifold

We will illustrate our reasoning with a familiar example of
a Calabi-Yau manifold -- a quintic hypersurface in ${\bf CP}^4$.
Since we want to be able to consider certain $(0,2)$ deformations
of the usual $(2,2)$ model, we formulate it as a $(0,2)$ linear
sigma model.
This can be done as follows
(see section 6.2 of  \phases).  We work in $(0,2)$ superspace
with fermionic coordinates $\theta^+$ and $\bar\theta^+$ and bosonic
coordinates $y^\alpha$, $\alpha=1,2$.
The goal is to incorporate the parameters determining the
size and complex structure of the manifold $X$ and those determining the
complex structure of the stable holomorphic
vector bundle $V$ as coupling constants of a linear sigma model which
at large radius reduces to the familiar nonlinear sigma model
on $(X,V)$ in the infrared.  In particular, the imaginary part of the Kahler
parameter, $r$, will arise as the coefficient
of a Fayet-Iliopolous $D$-term of a world-sheet $U(1)$ gauge group.
In terms of the gauge-covariant
superspace derivative ${\cal D}_+$ defined in (\phases, equations
6.2 and 6.4) the field strength superfield for the $U(1)$ gauge group is
$\Upsilon=[{\cal \bar D}_+, {\cal D}_0-{\cal D}_1]$.  In Wess-Zumino gauge it
becomes $\Upsilon=i\bar D_+V+\partial_-\bar D_+\bar\Psi$
where the superfields $\Psi=\theta^+\bar\theta^+(v_0+v_1)$ and
$V=v_0-v_1-2i\theta^+\bar\lambda_--2i\bar\theta^+\lambda_-
+2\theta^+\bar\theta^+D$ organize the gauge field $v_\mu$ and the
gaugino $\lambda_-$ into a $(0,2)$ gauge supermultiplet.
There is also a gauge neutral chiral multiplet
$\Sigma^\prime=\s+\sqrt{2}\theta^+\lambda_+
-i\theta^+\bar\theta^+(D_0+D_1)\s$
(which in a $(2,2)$ language would combine with $\Upsilon$ into
a twisted chiral superfield).

The complex structure and bundle parameters will come from
superpotential terms.  These will involve six chiral superfields
\eqn\bmult{\Phi^I=\phi^I+\sqrt{2}\theta^+\psi^{I}_+
-i\theta^+\bar\theta^+(D_0+D_1)\phi^I,\,\, I=0,\dots 5}
To construct the gauge-invariant superpotential of interest we will take
$\Phi^0=P$ to have charge $-5$ and the $\Phi^i=S^i,\,i=1\dots 5$ to have
charge $1$.  The bosonic components of these superfields are
called $p$ and $s^i$ respectively.
In addition, there are
six fermionic multiplets
\eqn\fmult{\Psi_-^{I}=\psi_-^I-\sqrt{2}\theta^+G^I-
i\theta^+\bar\theta^+(D_0+D_1)\psi_-^{I}-
2i\bar\theta^+Q^I\Sigma^\prime\Phi^I}
with the same
gauge charges $Q^I$ as the corresponding $\Phi^I$.
Here $G^I$ is an auxiliary field which gets integrated out in favor
of the (0,2) superpotential term $\bar J_I$ which is introduced below.
(In a $(2,2)$ language,
the $\Phi^I$ and $\Psi_-^{I}$ would combine into ordinary chiral superfields.)
The $\Psi_-$ obey a chirality condition
\eqn\hukak{\bar D_+\Psi_-^{I} = E^I}
where $E^I$ are holomorphic functions of the chiral superfields $\Phi^J$.
For $(0,2)$ models that arise as deformations of $(2,2)$ models,
\eqn\ukan{E^I=2iQ^I\Sigma^\prime\Phi^I.}
To actually write a solution of heterotic string theory, one would
also include additional degrees of freedom, such as free fermions,
to represent a left-moving $SO(10)\times E_8$ current algebra; we will
for the time being not need to make this explicit.

The Lagrangian consists of the
standard flat kinetic terms together with the following.
There is a $U(1)$ $D$-term with
coefficient $r$ and
and a $\theta$-term; together these can be written in the gauge-invariant form
\eqn\Dtheta{\eqalign{L_{D,\theta}
&={t\over 4}\int d^2y\,\,d\theta^+~
\Upsilon\biggr|_{\bar\theta^+=0} + h.c.
\cr
&=\int\,d^2y\biggl(-rD+{\theta\over{2\pi}}v_{01}\biggr)}}
\noindent where $t=ir+{\theta\over{2\pi}}$. There is also a $(0,2)$
superpotential
\eqn\sup{L_J=-{1\over \sqrt{2}}\int d^2y\,\,d\theta^+
\Psi_-^{I} J_I\biggr|_{\bar\theta^+=0} + h.c.}
Here
\eqn\defJ{J_0(\Phi^I)=G(S^i)}
where $G=G_{ijklm}S^iS^jS^kS^lS^m$ is the defining polynomial for the
quintic hypersurface, and
\eqn\defJII{J_i=F_{i,j_1j_2j_3j_4}S^{j_1}S^{j_2}S^{j_3}S^{j_4}P,~~i=1,\dots,
5}
is required to obey
\eqn\itobeys{S^iJ_i=5PJ_0=5PG(S^i).}
This ensures (using \ukan\ and the values of the charges)
that $\sum_I E^IJ_I=0$, a necessary condition for $(0,2)$ supersymmetry.
The model actually has $(2,2)$ supersymmetry if and only if
\eqn\hiun{J_i=P{\partial G\over \partial S^i}.}
It is convenient to set
\eqn\Jtilde{J_i\equiv P\tilde J_i(S^j)}

Departing from \hiun\ breaks $(2,2)$ supersymmetry
to $(0,2)$ and has the effect of perturbing the tangent bundle
of the quintic to a more general bundle $V$.  (Since the perturbed
bundle has rank three, the space-time gauge group is still $E_6\times
E_8$.)  This is seen as follows.
Massless left-moving
fermions satisfy $J^i\psi_{-,i}=0$ for $i=1,\dots,5$. A vector in $V$
can be described by five $v^i$, subject to the equivalence
$v^i\simeq v^i+\lambda s^i$, and satisfying $J_iv^i=0$.
These conditions are compatible on the hypersurface $G=0$ by virtue of
\itobeys.
If we decompose $\tilde J_i$ as
$\tilde J_i={\partial G\over{\partial s^i}}+G_i$ for some
quartics $G_i$ satisfying $G_i s^i=0$, then the
224 parameters in $G_i$ are the moduli that break $(2,2)$ down to $(0,2)$.

The model has a left-moving $U(1)$ symmetry, which we will call
$U(1)_L$ (it is actually part of a left-moving $E_6$ current algebra),
and a right-moving $U(1)$ $R$ symmetry, which we will call $U(1)_R$.
The charges carried by the various fields are as shown in Table 1.

\noindent {\bf \underbar{Table 1:  U(1) Charges}}
$$\vbox{\settabs 4 \columns
\+\underbar{Fields} &$\underline{(J_L, J_R)}$
&\underbar{Fields}&$\underline{(J_L, J_R)}$\cr
\+&&&\cr
\+$\psi^i_{+}$&(1/5, -4/5)&$\lambda_-$&(0,1)\cr
\+$\psi^i_{-}$&(-4/5, 1/5)&$\lambda_+$&(1, 0)\cr
\+$\psi^0_{+}$&(0, -1)&$\s$&(-1, 1)\cr
\+$\psi^0_{-}$&(-1, 0)&$s^i$&(1/5, 1/5)\cr
\+$p$&(0, 0)& & \cr}$$

The parameter space consists of $t$ and the $F_{i,j_1j_2j_3j_4}$.
If we subtract the 25 linear redefinitions of the $s^i$, these contain
326 independent parameters.  At large radius these
can be distinguished as one Kahler parameter,
101 complex structure deformations, and 224 deformations of the
holomorphic structure of $V$.
On the (2,2) locus $t$ and the 101 $G_{ijklm}$ are true moduli and
the left-moving supersymmetry provides an invariant distinction between
them:  $Q_-$ annihilates the mode conjugate to $t$ while
$\bar Q_-$ annihilates the modes conjugate to the $G$'s.
For the general (0,2) situation,  there is no
such algebraic distinction between the various modes,
and it is not clear whether theories parametrized by these variables
actually correspond in the infrared to conformal field theories;
exploring this point is precisely the goal of our investigation.

The parameter space we have just found is compact, or at least has
a natural compactification.  Consider first the $t$ variable.
Because of the invariance under $\theta\to\theta+2\pi, $ that is,
$t\to t+1$, the natural variable is $q=e^{2\pi i t}$.
It appears from this definition that $q$
runs over the whole complex plane except for the origin.  However,
it is natural to add two points: the point $q=0$ corresponds to the
infinite radius or field theory limit of the theory, and $q=\infty$ is the
Landau-Ginzburg point.  (These assertions can be seen semiclassically
\phases.) With these points included, the Kahler moduli
space is a copy of ${\bf CP}^1$, which is compact.

The other moduli can be treated as follows.
Rescaling the $F_{i,jklm}$ and the $G_{ijklm}$
by a common complex number $\lambda$ can be absorbed in a rescaling
of the $\Phi_i$ and the $\Lambda_i$ by $\lambda^{-{1\over 5}}$ up to a
change in the $D$ terms of the theory.  It is believed that the change
in the $D$ terms does not affect the infrared fixed point of the theory.
Accepting this and dividing by the overall scaling, the
parameter space of these modes is a complex projective space
${\bf CP}^{349}$; this is compact.  (We could also absorb
24 of these 349 parameters
in a linear redefinition
of the $s^i$ of determinant 1,
so there are only 325 physical deformations, as discussed
above.  Dividing just by scaling is more convenient and sufficient
to prove compactness.)  Of course, at some points in this ${\bf CP}^{349}$,
the model will become singular; this is part of what will be analyzed
later.

The bosonic potential energy of the theory (computed as in section 6.2 of
\phases) is
\eqn\bospot{\eqalign{U(\phi_I)
&={e^2\over 2}
\biggl(\sum_I Q_I|\phi_I|^2-r\biggr)^2+
\sum_I Q_I^2|\phi_I|^2|\s|^2+\sum_I|J_I|^2
\cr
&={e^2\over 2}
\biggl(\sum_i |s_i|^2-5|p|^2-r\biggr)^2+
|\s|^2\biggl(\sum_i|s_i|^2+25|p|^2\biggr)
+|G|^2+\sum_i|J_i|^2}}
The first term on the right hand side comes from integrating out
the auxiliary field $D$.
The theory can be studied semiclassically at large $|r|$;
for $r>>0$ we find the Calabi-Yau phase and for $r<<0$ the Landau-Ginzburg
phase at low energy.

More generally, we are interested in the infrared behavior of the theory
for general $t$ and $F_{i,j_1j_2j_3j_4}$.
Since we do not have direct access to
the infrared limit, we mostly restrict attention to computations which are
invariant under renormalization group flow.  This sector of the
theory is conveniently packaged in the various (quasi-)topological
twisted theories \topsig\egyang\mirtop\topgrav.  In these theories,
the stress tensor is shifted by
\eqn\Tshift{T_{\alpha\beta}\rightarrow T^\prime_{\alpha\beta}
=T_{\alpha\beta}-
{1\over 4}(\epsilon^\gamma{}_\alpha\partial_\gamma J_\beta
+\epsilon^\gamma{}_\beta\partial_\gamma J_\alpha)}

\noindent where $J$ is $J_R+J_L$ for the $A$-model, $J_R-J_L$ for the
$B$-model, and $J_R$ for the half-twisted model \foot{Actually we
are free to shift $J$
by the gauge current $\cal J$, which is $\{\bq,\dots\}$; we
will find it convenient to do so in section 4.}.  This has the effect
of shifting the spins of all fields by half their $J$ charge.  In
particular, $\bq$ becomes a scalar charge and plays the role of
a BRST operator, so that physical states of the twisted theory are
elements of $\bq$ cohomology.

In the $(0,2)$ case, the various twisted models are not topological
field theories, but they are still conformally invariant, as we will now
explain.
Because the action and measure are invariant under the symmetry
generated by $\bq$,
there is an interesting sector of the theory in which one considers
only operators annihilated by $\bq$.  Such
a correlation functions
vanishes if one of the operators is  $\bq$-trivial,
 $\alpha=\{\bq,\beta\}$ (and one does not meet anomalies from
surface terms).
We also have the relation $\Tr(T)=T_{+-}=\{\bq,\dots\}$,
a statement that holds in all three models because of the
underlying supersymmetry.
These properties ensure formally that all these models
are conformally invariant, since insertions of
$\Tr(T)$ vanish in correlation functions of physical observables.
Actually, these theories have a much stronger property: $T_{++}$
is similarly $\{\bar Q_+,\dots\}$, so that correlation functions
of $\bar Q_+$-invariant operators vary holomorphically in addition to
being conformally invariant.

We are interested in correlation functions of vertex
operators in the physical model.  The relation between the physical
and twisted models has to do with spectral flow \ttstar\agnt\bcov.
Shifting the spins as described above is realized in the path
integral by adding appropriate couplings of the fields to the
spin connection.
This will be implemented explicitly
for the computations of interest for us in section 3.
The space-time degrees of freedom undergo
spectral flow at the same time, accomplished by insertions of
space-time spin operators $S_\alpha$ or $S_{\dot\beta}$.

As explained
in \gepnotes, the $J_L$ and $J_R$ values classify the transformation
properties of the states under the space-time gauge and Lorentz groups.
The theory can be decomposed into the space-time (``external'') and
internal c=9 parts, so that $h^{tot}=h^{int}+h^{ext}$,
$q^{tot}=h^{int}+q^{ext}$ etc.  From the right-moving N=2 algebra
we have $h^{tot}_R\ge {1\over 2}|q_R|$, and for right-moving
NS states $h^{tot}_R={1\over 2}\Rightarrow q_R=\pm 1$.  A scalar in
space-time is invariant under the external U(1) so that
$q_R^{int}=q_R=\pm 1$.  Its fermion partner has $q_R=\pm{1\over 2}$.

The left U(1) is part of the spacetime gauge group.  We will be interested
in three types of three-point functions
involving the generations, antigenerations, and $E_6$ singlets:
${\bf{\bar{27}^3}}$,
${\bf 27^3}$, and
${\bf S^3}$.\foot{Note that in our notation, {\bf S} refers
to {\it any} $E_6$ singlet, not only those which are associated
to $H^1(End~V)$ at large radius.}
As just explained, twisting the model is equivalent to
inserting spectral flow generators.  In particular, the twisting by $J_R$
turns the two fermion vertex operators into boson vertex operators.
For the $A$ and $B$ models we also twist by $\pm J_L$, which takes
us from a spinor representation of $SO(10)$ with half-integral
$J_L$ (so that the {\it total} left-moving $U(1)$ charge is integral)
to a scalar or vector representation of $SO(10)$ with integral $J_L$.
In particular,
under $SO(10)\times U(1)$, the ${\bf \bar{27}}$ of $E_6$ decomposes as
${\bf \bar{27}}={\bf\bar{16}}_{1/2}\oplus {\bf 10_{-1}}\oplus
{\bf 1_2}$ and left spectral flow takes us from $ {\bf 10}_{-1}$ to
$ {\bf\bar{16}}_{1/2}$ to ${\bf 1}_2$.

%\eqn\twisting{\eqalign{<V_FV_BV_F>_{phys}
%&\propto <V^{q_1-{3\over 2}}_BV_BV^{q_2-{3\over 2}}_B>_{A, int}
%\cr
%&\propto <V^{q_1+{3\over 2}}_BV_BV^{q_2+{3\over 2}}_B>_{B, int}
%\cr
%&\propto <V^{q_1}_BV_BV^{q_2}_B>_{H, int}}}

%\noindent where $q_1$ and $q_2$ are internal $J_L$ charges, and where
%the external part of the amplitude has been evaluated in free field
%theory giving the proportionality factor.
More explicitly
the ${\bf \bar{27}}^3$ amplitude can be performed in the $A$-model
as follows:

\eqn\phystwstA{<V_F^{\bf \bar{16}_{1/2}}V_B^{{\bf 10}_{-1}}
V_F^{\bf \bar{16}_{1/2}}>_{phys}\propto
<V^{-1}_BV^{-1}_BV^{-1}_B>_{A,\, internal}}

\noindent where the proportionality factor is obtained by
evaluating the $SO(10)$ correlator in free
field theory, and where
the superscripts on the internal vertex operators indicate left U(1)
charges.
Similarly, for the ${\bf 27^3}$ amplitude we have the $B$-model computation

\eqn\phystwstB{<V_F^{\bf {16}_{-1/2}}V_B^{{\bf 10}_{1}}
V_F^{\bf {16}_{-1/2}}>_{phys}\propto
<V^{1}_BV^{1}_BV^{1}_B>_{B,\, internal}}

\noindent Finally, for the ${\bf S^3}$ amplitude we obtain a half-twisted
model computation:

\eqn\phystwstH{<V_F^0V_B^0
V_F^0>_{phys}=
<V^0_BV^0_BV^0_B>_{H, internal}}
We will introduce in section 4 the exact forms of the linear sigma model vertex
operators we will use in the various twisted computations.

\subsec{Location Of Singularities}

In order to pin down the space-time superpotential, we would
like to understand where this two-dimensional theory can become singular.
The basic strategy in analyzing singularities is as follows.
Suppose (this is a simplification) that for generic couplings
the bosonic potential \bospot\ had the property of going
to infinity as the fields go to infinity -- in any direction in field
space.  This growth of the potential at infinity would ensure the convergence
of the path integral.
Singularities could then only arise for parameters such that
(in some direction in field space) the growth of the potential
at infinity is lost.

Since we want to locate the singularities in the {\it quantum} theory,
what is really relevant here is the large field behavior of the quantum
effective potential.  Happily, because of the super-renormalizability of
the linear sigma model, the semiclassical approximation is valid
for large fields and (except for a slight shift due to
a one loop effect that will be noted in section 4)
the classical potential can be used in locating singularities.

A more serious problem in implementing this program is that
the assumption we made about the potential was too optimistic.
Looking at \bospot, we see that by setting $\Phi^I=0$, we can take
$\sigma$ to infinity for only a finite cost in action; the action
in this limit is $e^2Ar^2/2$, with $A$ the area of the surface.
(Actually, as we will discuss more thoroughly in section 4,
there is an additional term $e^2(A/2)(\theta/2\pi)^2$ coming from
the fact that in two dimensions a non-zero $\theta$ induces a stable background
electric field \coleman.)
This complicates the discussion, as we will analyze in section 4,
but only in a relatively mild way, roughly because in the infrared limit
(relevant to the conformal field theory we really wish to study)
$A$ is going to infinity.

However, it is clear that at some value of $t$ for $r$ near 0, the
barrier against going to large $\sigma$ might altogether vanish.
We will determine in section 4 precisely where this occurs and
show that a singularity actually does occur at the value of $t$ in question.

Singularities might also occur as functions of the $J_I$.
This will occur if the $J_I$ are such that the equations $J_I=0$
can be obeyed with $p\not= 0$ and some of the $s^i$ non-zero.
Then, taking $p$ and the appropriate $s^i$ to infinity (in such
a ratio as to ensure vanishing of the $D$ term),
one has vanishing classical potential, so one should expect a
divergence of the path integral.
We will study this region
in section 5, showing that it is isomorphic to the situation
near $t=0$ and does give a singularity.

Once we locate the singularities, we can determine which physical
quantities diverge there.  In particular, by showing that the ${\bf S}^3$
couplings have no singularities, we will be able to deduce -- since the
parameter spaces are compact -- that they vanish identically.

\newsec{Behavior near $r=\theta=0$}

We have just noted that the bosonic potential has a dangerous behavior
for $\phi^I=0,\sigma\to \infty$, and a still more dangerous behavior
when in addition $t$ is near 0.  We will in this section analyze
the behavior of the path integral in this region.  The analysis
is tractable because most of the fields (including the $\phi^I$)
have masses proportional to $|\sigma|$ for $\sigma\to\infty$, and the
light degrees of freedom become free in this region.

As discussed in the last section, the linear sigma model becomes singular at
$t=ir+\theta/2\pi=0$ due to the vanishing of the  potential of $\sigma$,
which produces an unbounded zero mode and a continuum of gapless excitations.
By studying the behavior of gauge-singlet
correlation functions near $t=0$, we can determine the order of
the pole in the spacetime superpotential.  Before doing this we will study the
well-known $\bt^3$ coupling in this region in order to check
our methods and provide a quick derivation of the simple pole in this
amplitude.  Then we will investigate whether the ${\bf S^3}$ correlators
can diverge at this locus.

To begin with we are interested in the Yukawa coupling
$<V^{{\bf \overline{27}}}_FV^{{\bf \overline{27}}}_B
V^{{\bf \overline{27}}}_F>$.  Specifically
we will study here the component
$<V^{{\bf \overline{16}}_{1/2}}_FV^{{\bf 10}_{-1}}_B
V^{{\bf \overline{16}}_{1/2}}_F>$.
This way the fermions are in the
(R, R) sector: the right-moving spinor ground state ensures
that the vertex operator describes a spacetime fermion, and the left-movers lie
in the Ramond sector so that the operator transforms in the spinor
(${\bf 16}$) representation of SO(10).
The boson vertex operator for the ${\bf 10_{-1}}$ component of the $\bt$ is
$\chi_j\sigma$ for $j=1,\dots,10$ where the $\chi_j$ are free left-moving
fermions in the vector representation of SO(10).
In the (2,2) model $\sigma$ is the lowest component of the twisted
chiral superfield $\Sigma=\sigma - i\sqrt{2}\theta^+\bar\lambda_+
- i\sqrt{2}\bar\theta^-\lambda_- +
\sqrt{2}\theta^+\bar\theta^-(D-iv_{01})+\dots$, so on the (2,2) locus
$\sigma$ is related as it should be by a left-moving supersymmetry
transformation to the Kahler modulus $t$.

It is straightforward to
check that $V_B^{{\bf 10}_{-1}}=\sigma$ is normalized in the standard fashion
at large radius, so that in particular $\langle \sigma^3\rangle$ has
no pole at large radius \grass.
(This will also help determine the normalization
of singlet vertex operators later.)
For $r>>0$ we have $\sum_i|s^i|^2=r$ and $p=0$.  The part of the
Lagrangian coupling $\s$ to the $\psi^i_\pm$ and $\bar\psi^i_\pm$ is
\eqn\siglag{{\cal L_\s}=\int\,d^2y\,\biggl(
r|\s|^2+\sqrt{2}\sum_i\bar\psi^i_+\bar\s\psi^i_-
+\sqrt{2}\sum_i\bar\psi^i_-\s\psi^i_+\biggr).}
In the $A$-model the fermions which have zero modes are the
three linear combinations of $\bar\psi^i_+$ and of $\psi^i_-$ which
are superpartners of the three combinations of $s^i$ which are
tangent to the manifold.
Since $r$ is very large, $\s$ is very
massive, and we can integrate it out using its equations of motion.
The $\bar\s$ equation gives
\eqn\sigform{\s=-{\sqrt{2}\over r}\bar\psi^i_+\psi^i_-.}
Consider the transformation laws of
$s^i$ and $\bar s^i$
under $\bq$ and $Q_-$:
\eqn\stransf{\eqalign{
&\{\bq,s^i\}=0,~~~\,\,\{\bq,\bar s^i\}=-\sqrt{2}\bar\psi^i_+;
\cr
&\{Q_-,s^i\}=\sqrt{2}\psi^i_-,~~~\,\,\{Q_-,\bar s^i\}=0.}}
These suggest that we interpret $\bq$ as the $\bar\partial$ operator
and $Q_-$ as the $\partial$ operator of the space of $s^i$.  Then
$\bar\psi^i_+$ is identified with $d\bar s^i$ and $\psi^i_-$ is
identified with $ds^i$.  Then from \sigform~ we see that
at large radius $\s$ becomes identified with a $(1,1)$ form,
normalized so (as the volume of $X$
is of order $r^3$ and $\sigma$ is of order $1/r$)
that $\int_X\,\s\wedge\s\wedge\s$
is independent of $r$.

To evaluate this correlator as it stands
we would need an expression for the covariant fermion vertex operators
in the linear sigma model.
It is easier to instead work in a ``twisted'' model
as described in the previous section.  In particular,

\eqn\fulltwist{<V^{\bf{\overline{16}_{1/2}}}_FV^{\bf{10_{-1}}}_B
V^{\bf{\overline{16}_{1/2}}}_F>_{phys}\propto
<\sigma\sigma\sigma>_A}

\noindent where the subscript $A$ refers to the analogue of the fully twisted
``$A$-model'', in which the spins of all fields are shifted by
${1\over 2}J=-{1\over 2}(J_L+J_R)+{1\over 5}Q$; here $Q$ is the
gauge charge which we can add for convenience as indicated
in the  footnote following equation \Tshift.  This allows us to compute
using only the simple internal bosonic vertex operator $\s$.  The
spin zero fields are now $\lambda_+$, $\bar\lambda_-$, $\psi_{+,0}$,
$\bar\psi^0_{-}$, $\psi^i_{-}$, $\bar\psi^i_{+}$, $s^i$, and $\s$.
($p$ is then twisted and will therefore later have no zero mode.)

\subsec{An Anomaly}

%--area anomaly

%--Hamiltonian framework, operator formalism, and normalization conditions for
%wavefunctions

Since the path integral has a dangerous behavior for $\sigma\to\infty$
with $\phi^I$ near zero, we first want to analyze the behavior in this
regime to make sure that the path integral does converge.  We will see
that it does converge but poorly enough that there are anomalies in some
formal assertions made in section 3.  Once these anomalies are understood,
we will be in a  position to proceed to more precise computations.

We will first compute the contribution of the large $\sigma$ region
to \fulltwist.  We can integrate out the
$\Phi_I$ multiplets, which have masses of order $\s$.
This will give an effective action for the light modes, $v_\mu$, $\lambda_-$,
$\bar\lambda_-$, $\s$, $\lambda_+$, and $\bar\lambda_+$.
which we will then study more carefully.
In fact, for large $\sigma$, $(2,2)$ supersymmetry is restored and the
light modes can be conveniently organized in a $(2,2)$ superfield
\eqn\sig{\eqalign{\Sigma
&=\sigma - i\sqrt{2}\theta^+\bar\lambda_+
- i\sqrt{2}\bar\theta^-\lambda_- +
\sqrt{2}\theta^+\bar\theta^-(D-iv_{01})
-i\bar\theta^-\theta^-(\partial_0-\partial_1)\s
\cr
&-i\bar\theta^+\theta^+(\partial_0+\partial_1)\s
+\sqrt{2}\bar\theta^-\theta^+\theta^-
(\partial_0-\partial_1)\bar\lambda_+
+\sqrt{2}\theta^+\theta^-\bar\theta_+
(\partial_0+\partial_1)\bar\lambda_-
\cr
&-\theta^+\bar\theta^-\theta^-\bar\theta^+
(\partial_0^2-\partial_1^2)\s}}

In fact, because of the large masses and the super-renormalizable nature
of the theory, the only relevant term is the one loop contribution that
comes by integrating out the $\phi^i$.  This gives (upon setting
$v_\mu\equiv 0$ in the last step, anticipating a stationary
point with this property):
\eqn\bosons{\eqalign{
&\int \prod_i d^2s^i d^2p \exp{-\int Ds^i D\bar s^i +DpD\bar p
+D(|s^i|^2-5|p|^2)+|\s|^2(|s^i|^2+25|p|^2)}
\cr
&=\biggl(\prod_i \det{1\over{-(\partial_\mu+iv_\mu)^2+D+|\s|^2}}\biggr)
\det{1\over{-(\partial_\mu-5iv_\mu)^2-5D+25|\s|^2}}
\cr
&=\exp\{\int\,{d^2k\over{(2\pi)^2}}\biggl[
-5\ln (k^2+D+|\s|^2)-\ln (k^2-5D+25|\s|^2)\biggr]\}}}

Performing the integral with constant $D$ and $\sigma$ and
exponentiating this back into the action, the effective
equation of motion for the auxiliary field $D$ (including also the classical
contribution $-Dr+D^2/2e^2$) is

\eqn\Deom{D-r+{5\over {2 \pi}}\ln 5 +
{5\over {4 \pi}}\ln ({{|\s|^2-{D\over 5}} \over {|\s|^2+D}})=0}

\noindent Expanding in $1/|\s|^2$ we find

\eqn\Dvac{D (1 - {3\over{2\pi |\s|^2}})={r-{5\over {2 \pi}}\ln 5}}

We can reproduce this equation of motion for $D$ in the following
(2,2) superspace Lagrangian (which was derived in the context of the
${\bf CP}^n$ model in \daddaetal):

\eqn\seff{S_{eff}=\int d^2y
\biggl(\int d^4\theta[{-({1\over {4e^2}})} \bar\Sigma\Sigma
 -{3\over {8\pi}}\ln\Sigma \ln\bar\Sigma]+
{it_R\over{2\sqrt{2}}}\int d\theta^+d\bar\theta^-\Sigma-
{i\bar t_R\over{2\sqrt{2}}}\int d\theta^-d\bar\theta^+\bar\Sigma\biggr)}
\noindent which is good for large $\s$.
Here the (2,2) twisted chiral superfield $\Sigma$ is given by
\sig, and $t_R$ is a renormalized Kahler parameter,
\eqn\geff{t_R=t-{5\over 2\pi}\ln 5.}

There are also $D$-independent terms in \bosons.  They largely cancel,
because of supersymmetry, against similar $D$-independent terms
in the fermion determinants.  (If one can set $D$ to zero, then
the region of large $\sigma$ and zero $\phi^I$ has unbroken supersymmetry.)
The cancellation is complete except for the ``constant'' modes of the
various fields, the zero modes of the kinetic energy.  The complex bosons
$s^i$ each have a constant
mode ({\it not} $p$, which is twisted by virtue of its gauge charge);
such a complex bosonic ``zero mode'' contributes a factor
of $1/\bar\sigma\sigma$ to the path integral.  For the fermions,
as we are in genus zero, only the components $\psi^0_{+}$,
$\bar\psi^0_-$, $\bar\psi^i_{+}$, and $\psi^i_-$
that are twisted to have spin zero have such ``zero modes''; each
pair $(\bar\psi^i_{+},\,\psi^i_-)$ contributes a factor of
$\bar\s$ and the pair $(\psi^0_{+},\,
\bar\psi^0_-)$ contributes a factor of $\s$.  Multiplying these factors,
the net $D$ independent contribution to the path integral is a factor
of $1/\s^4$.  This can be interpreted as the contribution
of an effective action

\eqn\schiral{\delta S_{eff}=\int d^2y\sqrt{g}{\hat R\over{4\pi}}
\bigl[-4\ln\s\bigr]}

\noindent This form of the effective action could also be predicted by studying
the underlying anomalies.

To evaluate

\eqn\effampl{\int d^2\s d\bar\lambda_-d\lambda_+ ~\sigma^3 ~
e^{-S_{eff}}}

\noindent we proceed as follows.
To absorb the fermion zero-modes here we bring down a factor
of $\int d^2y {2\sqrt{2}\over{\s\bar\s^2}}(D-iv_{01})
\bar\lambda_-\lambda_+$ from the component expansion of the
$\ln\Sigma \ln\bar\Sigma$ term in the action \seff.
We now must  recall  a subtlety alluded to earlier involving
the $\theta$-angle in two-dimensional electrodynamics: as
a function of $\theta$, the expectation value of $v_{01}$ is \coleman\
$-e^2\theta/2\pi$
(with $\theta$ understood to be in the range $|\theta|\leq \pi$).
Then \effampl\ becomes

\eqn\Adep{(-iA t_R)\int d^2\s {2\sqrt{2}\over{\s^2\bar\s^2}}
e^{-{{Ae^2|t_R|^2\over 2}}}}

\noindent with $A$ the area of the world-sheet.
So the integral converges for $\sigma\to\infty$ despite the poor behavior of
the classical potential in that region.
However, the convergence is relatively slow, and it is a familiar
story that when physical amplitudes are given by integrals that barely
converge, there often are anomalies in formal arguments about their
properties.  In this case, a possible anomaly suggests itself: despite
the formal argument in section 3 for conformal invariance, the above
formula certainly suggests that $\langle\sigma^3\rangle$ may be $A$ dependent.

Some more information is needed to check this, since
\Adep\ is only valid for large $\sigma$.
However, we will now demonstrate that there is indeed an anomaly
in the formal argument for conformal invariance.
%First, recall that reparameterization invariance allows us
%to gauge-fix the metric $g_{\mu\nu}$ to one of the form
%$g_{\mu\nu}=e^\phi g^{(0)}_{\mu\nu}$ for fixed metric $g^{(0)}_{\mu\nu}$.
%As suggested by \Adep, the result may, however, depend on the
%total area $A$.
We recall that the formal argument depends on writing the trace
of the stress-tensor as
\eqn\hucc{Tr(T)=\{\bq,B\},}
 where $B$ is a component of the supercurrent.
Given \hucc, one tries to prove conformal invariance as follows.
Under a conformal transformation $\delta g=\alpha g$,
the change in $\langle\sigma^3\rangle$ is proportional to
\eqn\conf{
<Tr(T(x))\s\s\s>=<\{\bq,B(x)\}\s\s\s>=0}
The last step really involves integration by parts on the $\sigma$
plane.

To make this systematic, we interpret the zero modes via differential
forms on the $\sigma$ plane.
Consider the transformation laws under the
supersymmetry transformation generated by $\bq$:  $\delta\sigma=0$ and
$\delta \bar\sigma = -i\sqrt{2}\bar\epsilon_-\lambda_+$.
We see that in the large $\sigma$ region,
it is natural to identify ${\bar Q_+}$ with the $\bar\partial$ operator
of the $\sigma$ plane and $\lambda_+$ with $d\bar\s$.
In particular, acting on the zero-modes of
the worldsheet fields,
\eqn\diffop{\bq=\sqrt{2}\lambda_+{\partial\over{\partial \bar\s}}+\dots}
where $\dots$ refers to terms which disappear when $t\to 0$.
Suppose we want to compute
$\langle \{\bar Q_+,\Lambda\} \rangle$ for some $\Lambda$.
In the $A$-model $\lambda_+$ and $\bar\lambda_-$ have spin zero.
After integrating out the nonzero modes we are left with
$\langle \{\bar Q_+,\Lambda\} \rangle$
as a function of the fermion zero modes $\lambda_+$ and $\bar\lambda_-$
as well as the bosonic zero modes $\s$ and $\bar\s$.
Integrating out the fermion zero-modes picks out the component
\def\llambda{\eta}
$\llambda$ of $\Lambda$ which multiplies the fermion zero-mode
$\bar\lambda_-$.  Then we are left with
an integral
\eqn\yip{\langle \{\bar Q_+,\llambda\}\rangle
=\int_Ud^2\sigma~{\partial\llambda\over{\partial\bar\s}}
=\int_{\partial U}d\sigma ~\llambda,}
where $U$ is the $\sigma$ plane and $\partial U$ is a circle at infinity.
%and to interpret ???? as $d\bar \sigma$.
%Likewise, we identify ??? as $d\sigma$.
%Integration $\int d^2\sigma d?? d?? f(\sigma,??)$ picks out the
%component of $f$ proportional to both $??$ and $??$, that is, the
%component which is a $(1,1)$ form.  Suppose we want to compute
%$\langle \{\bar Q_+,\Lambda\} \rangle$ for some $\Lambda$.  By integrating
%out the non-zero modes, we express $\Lambda$ as a function $\lambda$ of
%the zero modes -- that is, a differential form on the $\sigma$ plane.
%Then $\{\bar Q_+,\Lambda\}$ corresponds to the differential form
%$\bar\partial \lambda $.  Then we get
%\eqn\yip{\langle \{\bar Q_+,\lambda\}\rangle=\int_U\bar \partial\lambda
%=\int_{\bar\partial U}\lambda,}
We can schematically write $\int_{\bar \partial U}\llambda
=\langle \Lambda \rangle'$ where $\langle \dots\rangle'$ is
a modified correlation function in which one suppresses the zero mode
of the fermion field $\bar\lambda_-$ and integrates
only over a circle at infinity
in the $\sigma$ plane.
With this understood, \conf\ becomes
\eqn\polyni{\langle T\sigma^3\rangle=\langle b\sigma^3\rangle'.}
where $b$ is the component of $B$ which multiplies $\bar\lambda_-$.
Thus, we have a precise framework for detecting a possible anomaly.

{}From the effective action, one finds that
\eqn\trt{\eqalign{Tr(T)
&={-3\sqrt{2}\,\bar\lambda_-\lambda_+\over{4\pi\s\bar\s^2}}(D-iv_{01})
+{3\sqrt{2}\,\bar\lambda_+\lambda_-\over{4\pi\bar\s\s^2}}(D+iv_{01})
\cr
&+{3\bar\lambda_+\lambda_-\bar\lambda_-\lambda_+\over{2\pi |\s|^4}}
-rD+{D^2\over 2}(1-{3\over{2\pi |\s|^2}})}.}
Hence using the supersymmetry transformation laws
$\{\bq,\bar\s\}=i\lambda_+\sqrt{2}$,
$\{\bq,\bar\lambda_-\}=-i(D+iv_{01})$, and
$\{\bq,D\}=-\partial_+\lambda_-$ one has
\eqn\trivT{B=(D-iv_{01}){\bar\lambda_-\over 2}
\bigl(1-{3\over{2\pi |\s|^2}}\bigr)
+{3i\over{2\pi\sqrt{2}}}
{\bar\lambda_-\bar\lambda_+\lambda_-\over{\s^2\bar\s}}.}
To find the component $b$ of $B$ multiplying $\bar\lambda_-$ for large
$\sigma$ is very
easy: one replaces $D-iv_{01}$ by its expectation value $-i t_R$, and
and one can discard the terms proportional
to $1/|\sigma|^2$ or $\bar\lambda_+\lambda_-$ because they are
negligible for large $\sigma$.  So one has simply
$b=-i t_R /2$.  In computing $\langle b\sigma^3\rangle'$
we also need the behavior of the effective action in the large
$\sigma$ region: as derived above, this is
\eqn\puriv{\exp (-S_{eff})\sim \sigma^{-4}\exp(-{Ae^2|t_R|^2\over 2}).}
The factor of $\sigma^{-4}$ is from the ``zero modes'' analyzed above,
while the factor of $\exp(-{Ae^2|t_R|^2\over 2})$
comes from the behavior of the
effective potential in the large $\sigma$ region
including the effects of the $D$ term and the $\theta$ angle as
explained above.

Putting this together, we get
\eqn\anom{\eqalign{
<b\s\s\s>'
&=\int_{\partial U} {d\sigma\over \sigma}{1\over 2}(-i t_R) \exp\left(
{-Ae^2|t_R|^2\over 2}\right)
\cr
&= \pi t_R \exp\left(-{{Ae^2|t_R|^2\over 2}}\right)}}
So the surface term is not zero and the naive conformal invariance
is violated.

On ${\bf S}^2$, any two metrics are conformally equivalent, so the metric
dependence of the correlation function can be determined by integrating the
conformal anomaly.  As the anomaly formula \anom\ depends only on the
area and not on other details of the metric, the correlation function has
the same property.

\subsec{Hamiltonian Formalism}

Because of this anomaly in conformal invariance of the twisted model, we
must proceed with care
to extract the correct physics.
We cannot simply compute a correlation function on ${\bf S}^2$ with
any given finite area: the result will depend on the area.  It is clear,
though, what we need to do to get the right result.
If it is true that the linear sigma model
flows in the infrared to the conformal field theory that we really
want, then by taking $A\to\infty$ in the linear model, we will get the
desired results.

The precise way of taking $A\to\infty$ does not matter, since
as we have just seen the result only depends on the total $A$.
The most obvious way to go to large $A$ is to scale up the metric
$g$ of the surface by $g\to e^\alpha g$ with $\alpha $ a large constant.
But in that limit, the linear model is not tractable: trying to compute
in that region is like trying to explicitly understand the renormalization
group flow of the model in the infrared.

An alternative way to go to large $A$ is to expand the sphere in only
one direction, so that it becomes a very long cigar, of circumference
$2\pi$ and length $A/2\pi$.
We insert one copy of $\sigma$ at the left end of the cigar, one at the
right end, and one in the middle.
In this limit we reduce to a Hamiltonian
framework: the path integral on half a cigar with $\sigma$ inserted
at the tip gives a quantum state $\Psi$, and the correlation function
we want is $\langle\Psi|\sigma|\Psi\rangle$.
We can assume  $\Psi$ is a state of zero energy, since in the limit
that the cigar is very long, other components are exponentially damped.

To find possible singularities, we only need to know how $\Psi$ behaves
for large $\sigma$.
A zero energy state must obey
\eqn\Qkill{\bq \Psi = Q_+ \Psi =0.}
This together with the quantum numbers determines the large $\sigma$
behavior of $\Psi$ up to a normalization constant.
To find singularities, we only need to know the normalization constant
up to a finite factor. This is given as follows.
In computing $\Psi(\sigma,\bar\sigma,\bar\lambda_-,\lambda_+)$,
one performs a path integral on a half-cigar with the boundary values
fixed.  Fixing the boundary values eliminates the zero modes, so
there is no singularity in the wave-function; $\Psi(\sigma,\dots)$
has a pointwise limit even when the bare couplings are taken to a value
at which the theory becomes singular.  Singularities in the correlation
functions will have to arise because giving $\Psi(\sigma,\dots)$
a limit as couplings approach a dangerous value with $\sigma$ fixed
will result in $\Psi $ being unnormalizable (or so weakly normalizable
that correlation functions diverge).

\subsec{The ${\bf{\overline {27}}^3}$ Coupling}

%--computation of 16-16-10 component of ${\overline {27}}^3$

%--agreement with mirror symmetry result but from direct LSM computation

In the large $\s$ regime, we can set the $\Phi^I=0$ since their
masses are of order $\s$.  In this regime the model can
be analyzed semiclassically (as done in some detail above) and
we are free to work in the physical model where the fields
have canonical spins.   Taking this option, we have the
following explicit expressions for the supercharges:

\eqn\Qfldth{\eqalign{
&\bq=\int dx^1
\bigl({\sqrt{2}}\lambda_+\partial_+\s -\lambda_-t_R\bigr)
\cr
& Q_+=\int dx^1
\bigl({\sqrt{2}}\bar\lambda_+\partial_+\bar\s -\bar\lambda_-\bar t_R\bigr)}}

\noindent Truncating to zero modes and canonically quantizing, this becomes

\eqn\Qqm{\eqalign{
&\bq=
\sqrt{2}\lambda_+i{\partial\over{\partial\bar\s}}-\lambda_-t_R
\cr
& Q_+=
\sqrt{2}\bar\lambda_+i{\partial\over{\partial\s}}-\bar\lambda_-\bar t_R}}

The quantum wavefunction will depend on half the fermion variables;
we take $\Psi=\Psi(\s,\bar\s,\bar\lambda_-,\lambda_+)$.  The canonically
conjugate fermions act by differentiation, as follows from the
canonical anticommutation relations
\eqn\can{\eqalign{
&\{\bar\lambda_-,\lambda_-\}=1,\,\,\{\bar\lambda_+,\lambda_+\}=1;
\cr
&\{\lambda_-,\lambda_-\}=0=\{\bar\lambda_-,\bar\lambda_-\};
\cr
&\{\lambda_+,\lambda_+\}=0=\{\bar\lambda_+,\bar\lambda_+\}.}}

\noindent We will look for a bosonic ground state wavefunction, killed
by \Qqm, of the form

\eqn\wfexp{\Psi(\sigma,\bar\sigma,\bar\lambda_-,\lambda_+)=
f_1(\s,\bar\s)+f_2(\s,\bar\s)\lambda_+\bar\lambda_-}

\noindent Then the amplitude is

\eqn\hamamp{\int d^2\s d\lambda_+ d\bar\lambda_- \Psi\Psi\s
=\int d^2\s f_1(\s,\bar\s)
\s f_2(\s,\bar\s)}

\noindent Imposing \Qkill~ we find

\eqn\lineqs{\eqalign{
&\bq\Psi=0\Rightarrow i\sqrt{2}\bar\partial f_1+t_R f_2=0
\cr
&Q_+\Psi=0\Rightarrow i\sqrt{2}\partial f_2-\bar t_Rf_1=0}}

\noindent Combining these equations gives

\eqn\quadeq{(\partial\bar\partial-{|t_R|^2\over 2})f_1=0}

For each partial wave sector, this equation has two solutions; for generic
$t_R$,  one
of these vanishes exponentially at infinity and so is normalizable.
(It is not obvious from this approximation that the solution that
is normalizable for large $\sigma$ is also regular near $\sigma=0$,
but that follows, for instance, from the representation of $\Psi$
by a path integral on the half-cigar.)
A singularity in $\langle\sigma^3\rangle$ can only come when
the ``regular'' solution loses its normalizability, and this will
only be for $t_R=0$. (That $t_R=0$  is
the precise location of the singularity can also be checked in other
ways \ref\morrison{D. Morrison and R. Plesser, ``Summing the
Instantons: Quantum Cohomology and Mirror Symmetry in Toric
Varieties'', DUKE-TH-94-78, IASSNS-HEP-94/82, hepth/94122236.}.)

It remains to determine whether there really is a singularity
at $t_R=0$ and to compute its nature.
Separating variables, we can write the solution in the
form $f_1=N(t_R,\bar t_R)({\bar\s\over\s})^\nu\tilde f_1(|\s||t_R|)$ where
$N$ will be determined by the normalization conditions.

As explained above, the normalization condition is that
for fixed $\s$, as $t_R\rightarrow 0$, $f_1$ (and $f_2$) must be regular.
At $t_R=0$, \lineqs~ tells us that  $\bar\partial f_1=0$.  This combined with
the requirement of $U(1) $ charge conservation in \hamamp~ implies
$f_1(\sigma,t_R=0)
={1\over\s}=({\bar\s\over\s})^{1\over 2}{1\over{\as\at}}N(t_R)$
from which we learn that $N(t_R)=\at$ for small $t_R$.
Similarly for $t_R=0$ we find $\partial f_2=0$.  Then since
$\lambda_+\bar\lambda_-$ has the same left and right U(1) charges $(1,-1)$
as $f_1$, we learn that $f_2$ is constant for small $t_R$.

Now that $N$ is known, we can determine the behavior for large $\sigma$
with fixed $t_R$. The solution of the radial equation
is

\eqn\asymp{f_1=\at ({\bar\s\over\s})^{1\over 2}
{e^{-\sqrt{2}\as\at}\over{\as^{1\over 2}\at^{1\over 2}}}+\dots}

\noindent where we have inserted the normalization factor
$N(t_R)=\at$ and where
$\dots$ refers to the subleading terms in the asymptotic
expansion of the solution. (The solution is in fact the Bessel function
$K_\nu(\sqrt{2}\as\at)$ with $\nu={1\over 2}$).  From \lineqs~ we now find

\eqn\ftwo{f_2=i{\sqrt{2}\over t_R}\bar\partial f_1=
-i{\at\over t_R}\bigl[{1\over{2\sqrt{2}\as}}-\at
\bigr]{e^{-\sqrt{2}\as\at_R}\over
{\as^{1\over 2}\at^{1\over 2}}}}

\noindent So the amplitude becomes

\eqn\hamresult{\int d^2\s \s f_1 f_2=
\int d^2\s\s\at({\bar\s\over\s})^{1\over 2}
{e^{-2\sqrt{2}\as\at}\over{\at\as}}{i\at\over t_R}
\biggl({1\over{2\as}}+\at\biggr)
\sim {1\over t_R}}

\noindent This agrees with the simple pole in the ${\bf\bar{27}}^3$
coupling discovered using mirror symmetry by Candelas, de la Ossa, Green,
and Parkes \cdgp.

\subsec{${\bf S^3}$ Couplings}

%--review of half-twisted model, including table of the shifted spins
%and the integration measure, and including spectral flow from
%$\int J \bar\omega$

%\noindent {\bf Argument II}:  twisting of $\sigma$ $\Rightarrow$ no zero mode
%$\Rightarrow$ no poles

To compute the dependence on gauge-singlet
fields of the space-time superpotential $W$,
we study three-point functions of the singlets. This allows us
to read off three covariant derivatives of $W$.  This follows from the
fact that there is in the effective supergravity action a term
${1\over 2}e^{K/2}D_iD_jW\chi^i\chi^j$, where the $\chi$'s are matter
fermions in spacetime and $K$ is the spacetime Kahler potential.
We will show that this coupling is zero by showing that keeping
fixed all variables except the Kahler modulus $t$, the ${\bf S}^3$
coupling has no pole.

The only possible pole would be at $t_R=0$,
where the model is singular; we must show that the ${\bf S}^3$ couplings
have no pole there.\foot{The ${\bf S}^3$ couplings vanish
for $t=\infty$
because in this model the singlets are true moduli in the large radius,
field theory limit. Of course, a holomorphic function on a compact
complex manifold with a zero and no pole is identically zero.
As noted in the introduction, because the superpotential
is a section of a line bundle of negative curvature, its vanishing
follows from absence of poles even if one does not know of a zero.}

In this computation, it
is simplest to use the half-twisted model, in which the spins of
the fields are shifted by $-J_R/2$ (plus possibly a shift proportional
to the gauge charge $Q$ chosen for convenience).
This leads to the following spins
for the worldsheet fields (final spins of zero are written in bold-face
as those correspond to zero modes that are particularly important
in what follows):

\noindent {\bf \underbar{Table 2:  Spins in the half-twisted model}}
$$\vbox{\settabs 4 \columns
\+\underbar{Fields} &$\underline{ -({J_R\over 2}-{Q\over{10}})}$
&\underbar{Original Spin}&\underbar{Final Spin}\cr
\+&&&\cr
\+$\psi^i_{+}, \bar\psi^i_{+}$&(1/2, -1/2)&(1/2, 1/2)&(1,{\bf 0})\cr
\+$\bar\psi^i_{-}, \psi^i_{-}$&(0, 0)&(-1/2, -1/2)&(-1/2, -1/2)\cr
\+$\psi^0_{+}, \bar\psi^0_{+}$&(0, 0)&(1/2, 1/2)&(1/2, 1/2)\cr
\+$\bar\psi^0_{-}, \psi^0_{-}$&(1/2, -1/2)&(-1/2, -1/2)&({\bf 0}, -1)\cr
\+$\lambda_-, \bar\lambda_-$&(-1/2, 1/2)&(-1/2, -1/2)&(-1, {\bf 0})\cr
\+$\bar\lambda_+, \lambda_+$&(0, 0)&(1/2, 1/2)&(1/2, 1/2)\cr
\+$\s, \bar\s$&(-1/2, 1/2)&(0, 0)&(-1/2, 1/2)\cr
\+$s^i, \bar s^i$&(0, 0)&(0, 0)&({\bf 0, 0})\cr
\+$p, \bar p$&(-1/2, 1/2)&(0, 0)&(-1/2, 1/2)\cr}$$

\noindent With this prescription
we obtain the gauge-invariant zero-mode integration measure

\eqn\htmeasure{d\mu_h=d\bar\lambda_-d\bar\psi^0_{-}
\prod_id\bar\psi^i_{+}\prod_id^2 s^i}

\noindent which has left and right $U(1)$ charges $(0, -3)$.  The
${\bf S^3}$ correlation functions contain three vertex operators of
charge (0,1) ($q_L$=0 because these are gauge-singlets and $q_R$=1
for bosonic vertex operators).  So in the half-twisted model we can
compute ${\bf S^3}$ correlators with no extra spectral flow insertions
required.

The vertex operators for singlets can be identified by their
U(1) charges $(q_L,\,q_R)=(0,1)$, their correspondence with
parameters determining the size and shape of the manifold
and vector bundle,  and their relation to the (2,2) moduli on the
(2,2) locus.  The spacetime mode $R$ corresponding to the
Kahler parameter $t$ comes from the worldsheet gauge multiplet.
Its vertex operator must be annihilated by $\bq$ but should
be related by $Q_+$ to the combination $(D-iv_{01})$ which
occurs in the worldsheet action with coefficient $t$.  On the
(2,2) locus it is also related by a {\it left}-moving
supersymmetry generator to the vertex operator for the ${\bf 10_{-1}}$
component of the ${\bf\bar{27}}$,
which we determined above to be $\s$. These properties uniquely fix
the vertex operator for $R$:
\eqn\Rvert{V^R_B=\lambda_-}

The spacetime modes corresponding to the complex structure
and bundle parameters should be associated with sets of five quartic
polynomials $H^{(4)}_i(s^j)$ subject to one quintic relation.
They must be annihilated by $\bq$.  On the
$(2,2)$ locus the ones corresponding to complex structure deformations
must be related by a left-moving supersymmetry generator to
the vertex operator for the ${\bf 10_1}$ component of the
${\bf 27}$ vertex operator.  The latter are given by quintic
polynomials in the $s^j$ (these have left and right $U(1)$ charges
$(1,1)$ and are annihilated by $\bq$):
\eqn\tvert{V^{\bf 10_1}_B=pL^{(5)}(s^j)}
We have normalized this vertex operator to be independent of $r$,
ensuring that the ${\bf 27^3}$ coupling is constant on the Kahler
moduli space on the $(2,2)$ locus \distgreene.

Now let us discuss the vertex operators for $E_6$ singlets coming
from complex structure and bundle deformations.
Consider the vertex operator
\eqn\Svert{V_S=5pH^{(4)}_i(s^j)\psi^i_{-}+L^{(5)}(s^i)\psi^0_{-}}
where $H^{(4)}$ and $L^{(5)}$ are quartic and quintic polynomials,
respectively.  The condition for the operator to be $\bar Q_+$-invariant
is  $s^iH^{(4)}_i=L^{(5)}$.
101 such operators obey $H^{(4)}_i={\partial L^{(5)}\over{\partial s^i}}$
and are related near field theory to the deformation of the complex
structure of the manifold.
On the (2,2) locus, these operators are produced by acting on
the operators in \tvert\
with $Q_-$.
We normalized the vertex operator \tvert~ to be independent
of $r$.  Then the normalization of the complex structure vertex operators
discussed here follows from their relation to the ${\bf 27}$ on
the $(2,2)$ locus.
In any case it is clear that the vertex operators for the
complex structure and bundle deformations should be independent
of $r$ from the fact that the terms in the linear sigma model
action involving the parameters
$J_{i,j_1j_2j_3j_4}$ are decoupled from the term proportional to $r$.

In the ${\bf {\overline {27}}^3}$ coupling computed in the last subsection,
we found a simple pole singularity for $t\rightarrow 0$ due to the
unbounded zero-mode of $\sigma$.
{}From the above table of half-twisted spins, we
immediately see that no singularity is possible for the ${\bf S^3}$
correlator, for the simple reason that $\s$ has acquired a spin!  There
is now no bosonic zero-mode (recall that in this regime the $s_i$ have
huge masses of order $\s$) and hence no pole.  As discussed
in section 1, the spacetime superpotential must diverge on a locus
of codimension one in the moduli space if it is not to vanish identically.
We see here that this cannot happen, and that therefore the superpotential
is flat and the singlets are good moduli.

\subsec{${\bf {\overline {27}}^3}$ Revisited}

%--puzzle of the 10-10-1 coupling

%--puzzle of the 1 vertex operator

%--emergence of $\sigma$ zero-mode from $\int J \bar\omega$ term in action
%coming from the 1 vertex operator

This argument may seem a little too quick, for the following reasons:

(1) We obtained a simple pole for the the
${\bf 16-16-10}$ component of the ${\bf \bar{27}^3}$
coupling by doing the computation in the physical model in section 4.3.
We then argued that the ${\bf S^3}$ correlator is naturally done in
the half-twisted model and thus cannot have a similar
pole because $\s$ is twisted.  How do we reproduce the pole in
the ${\bf 16-16-10}$ coupling in the half-twisted model?

(2) By $E_6$ symmetry we should be obtain the same
simple pole by computing the ${\bf 10-10-1}$ component of
the ${\bf \bar{27}^3}$ coupling.

We will explain presently the resolution to problem (1).
There the spin
operator insertions needed to perform the ${\bf 16-16-10}$ computation
in the half-twisted model make up for the twisting of $\s$ and
we can recover the simple pole.
As for case (2), we find ourselves unable to find a natural
representative for the vertex operator for the ${\bf 1_2}$ mode in
the ${\bf 10-10-1}$ computation.
(This may be related to the fact that this mode appears in a twisted sector at
the Landau-Ginzburg  point.  It is certainly related to the fact
that the $(0,2)$ $E_6$ models that we are studying do not have
deformations to $(0,2)$ linear sigma models with $SO(10)$ gauge group.
\foot{That in fact is why, in order to study the $SO(10)$ models,
Distler and Kachru modified  the model to eliminate the
$\sigma$ field \dkfirst; for reasons that are still not entirely
clear, this did not give a model with the right properties.})
This leaves us without a satisfying
answer to question (2) within the linear sigma model.
However, we have no such difficulties
with the singlet coupling ${\bf S^3}$. (The singlet vertex operators and
their normalizations are conveniently determined in the linear
sigma model as described in the previous section.)  Therefore
we remain convinced of the absence of a pole in the singlet
coupling and the flatness of the spacetime superpotential.

To answer the first question, we will
translate the above computation of the {\bf 16-16-10} amplitude
into a half-twisted computation, working with $\s$ and $\lambda$ twisted
as in table 2.  But now we must explicitly insert the left spectral-flow
generators (i.e. internal part of the gaugino vertex operator for
the $U(1)$ which combines with $SO(10)$ to form a maximal subgroup of $E_6$)
to put the fermion states in the ${\bf \bar{16}_{1/2}}$ (spinor)
representation of $ SO(10)$.  It is not clear how to represent these insertions
by linear sigma model operators, but luckily there is another option:
exponentiate them, explicitly adding to the action the extra
couplings of the fields to the
spin connection as described in section 3.

In the half-twisted model we would compute

\eqn\half{<\biggl(\oint_{C_{z_1}}dw\Sigma(w)
V_B^{{\bf 10_{-1}}}(z_1,\bar z_1)\biggr)V_B^{{\bf 10_{-1}}}(z_2,\bar z_2)
\biggl(\oint_{C_{z_3}}dv\Sigma(v)
V_B^{{\bf 10_{-1}}}(z_3,\bar z_3)\biggr)>_{half}}

\noindent where $\Sigma$ is the left-moving analogue of the holomorphic
part of the
spacetime supersymmetry generator; in the CFT its internal part would be
given by $e^{i{\sqrt{3}\over 2}H}$ where H is the bosonization of
the left-moving U(1) current.  In the CFT we could
achieve these insertions by adding to the action a term
proportional to $\int H \hat R$
where $\hat R$ is the worldsheet curvature which has delta-function
support on the insertions.  Here we are working on an infinitely elongated
sphere, so that the bulk of the worldsheet is a flat cylinder and the
curvature is pushed out to the ends at infinity.
So we will take a step function for the spin connection, which
is constant everywhere except at the insertions.
On the bulk of the worldsheet, which
is a flat cylinder, this constant spin connection contribution
shifts the boundary conditions of the fields around the cylinder
since $\oint\omega_1 dx^1=1$.
The resulting contribution to
the curvature $\hat R$ is $\sim \delta(\tau-T)$, localizing the insertion
on a contour $\tau=T$ surrounding the operator on the end of the
cylinder.

To implement this plan in the Hamiltonian framework we must use
the following twisted mode expansions, which follow from the spins
in the half-twisted model (table 2):

\eqn\bmodes{
\eqalign{
&\s=a(\tau)e^{i{x\over 2}}+b(\tau)e^{-i{x\over 2}}+\dots
\cr
&\bar\s=\bar a(\tau)e^{-i{x\over 2}}+\bar b(\tau)e^{i{x\over 2}}+\dots}}
and
\eqn\fmodes{\eqalign{
&\lambda_+=\tilde\alpha(\tau)e^{i{x\over 2}}
+\tilde\rho(\tau)e^{-i{x\over 2}}+\dots
\cr
&\bar\lambda_+=\alpha(\tau)e^{-i{x\over 2}}
+\rho(\tau)e^{i{x\over 2}}+\dots
\cr
&\lambda_-=\gamma(\tau)e^{-ix}+\xi(\tau)+\dots
\cr
&\bar\lambda_-=\bar\lambda_-(\tau)+\dots}}
Our states will be killed by the supersymmetry generators

\eqn\halfQ{\eqalign{
&\bq=\int dx\bigl({\sqrt{2}}\lambda_+(\partial_+
-{i\over 2}\omega_+\bigr)\s
-t\lambda_-)
\cr
&Q_+=\int dx\bigl({\sqrt{2}}\bar\lambda_+(\partial_+
+{i\over 2}\omega_+)\bar\s
-\bar t\bar\lambda_-\bigr)}}
Note that to implement the insertion of the left spectral
flow operators,
we have explicitly included the spin connection terms.  This results
in the shift $\partial_+\rightarrow\partial_+ +i\omega_+{q_L\over 2}$
where $\omega_+$ is the + component of the spin connection.
For the first component of the spin connection we take

\eqn\spinconn{\omega_1=
\lim_{T\to\infty}\cases{
\theta(\tau+T),\tau < T;
\cr
\theta(T-\tau), \tau > -T.\cr}}
where $\theta (y)$ is the standard step function.
($\theta(y)=1$ for $y\ge 0$;
$\theta(y)=0$ for $y<0$.)
In terms of canonical variables \halfQ~ becomes

\eqn\halfQqm{\eqalign{
&\bq=\sqrt{2}\tilde\rho\bigl(i{\partial\over{\partial\bar a}}
+{ia\over 2}(1-\omega_1)\bigr)
+\sqrt{2}\tilde\alpha\bigl(i{\partial\over{\partial\bar b}}
-{ib\over 2}(1+\omega_1)\bigr)
-t\xi
\cr
&Q_+=\sqrt{2}\rho\bigl(i{\partial\over{\partial a}}
-{i\bar a\over 2}(1-\omega_1)\bigr)
+\sqrt{2}\alpha\bigl(i{\partial\over{\partial b}}
+{i\bar b\over 2}(1+\omega_1)\bigr)-\bar t\bar\lambda_-}}

\noindent With the choice \spinconn~ of spin connection we effectively
have a ``sudden'' pertubation turned on at $\tau=\pm T$.  The Hamiltonian
for $|\tau|>|T|$ has a quadradic potential for the modes $a$ and $b$, but
in the bulk ($|\tau|<|T|$) we have a constant
$\omega_1=1$ so the potential for $a$ is turned off:
a zero-mode for $a$ has
arisen because of the coupling to the spin connection (said differently,
$\s$ is not constant but is effectively covariantly constant, which
is what counts in computing the energy once the spin connection coupling
is included).  So we can reproduce the pole in the ${\bf 16-16-10}$
coupling in the half twisted model.

\newsec{Behavior for singular $F_{i,j_1j_2j_3j_4}$}
\subsec{Linear Sigma Model Analysis}
%--generic situation: one unbounded bosonic zero-mode, isomorphic
%to large $\sigma$ region for $r=\theta=0$

%--Distler/Kachru models with no $\sigma$ field:  still have potential
%singularities for special F's, which we can deal with here.

We now discuss what
happens at the other type of singularity discussed in section 2,
where the complex structure of the manifold and/or the gauge bundle
degenerates, keeping fixed the Kahler parameter $t=ir+\theta/2\pi$.

Since the part of the bosonic potential that comes from the
superpotential is
\eqn\bosup{ |p|^2\sum_i|\tilde J_i|^2 +|G|^2,}
for the charged fields to be able to go to infinity with finite cost in energy
requires that there be a non-trivial solution of
\eqn\rifpo{\tilde J_i=G = 0.}
(The $D$ terms are such for some $s^i$ to go to infinity at finite cost
in energy, $p$ must also, and vice-versa.)
As $5G=\sum_is^i\tilde J_i$, there are only five independent equations in
\rifpo; on the other hand, because the $\tilde J_i$ are homogeneous functions
of the $s^i$, in looking for a non-trivial function there are only
four independent variables, the ratios $s^i/s^5$ with $i\le 4$.  Hence
\rifpo\ generically has no solutions.  In looking for a pole in the
space-time superpotential, we need to consider the behavior when
one parameter is varied in the $\tilde J_i$ -- as poles, if they exist,
arise in codimension one on the parameter space.  By varying this
one parameter in the $\tilde J_i$ together with the four $s^i/s^5$, we have
five parameters in all, so generically we can expect finitely many
isolated solutions to the five equations $\tilde J_i=0$.

This means that the situation that we need to consider is that where,
for an exceptional value of the parameters, there is on ${\bf CP}^4$
an isolated solution of $\tilde J_i=0$.  This singularity
will occur on a generic point
$w$ on ${\bf CP}^4$ -- to impose a restriction on where on ${\bf CP}^4$
the solution will arise, one would need to adjust more than one parameter
in the $\tilde J_i$.    We may as well assume that the solution is at
$s^i=(0,0,0,0,1)$.  Also, when just one parameter in the $\tilde J_i$ is
adjusted, the Jacobian $\det(\partial_i \tilde J_j)$  will be
generically non-zero at $w$.

As practice, let us first show that the
${\bf 27}^3$ Yukawa couplings have a pole in this situation
(on the $(2,2)$ locus this is a standard result due to \candelas\ and
\toplg.)
In the generic situation described above, the only charged fields
that can become large are $s_5$ and $p$.  They must become
large in proportion because of the $D$ terms, and in the region
$s^5\to\infty$, $\sigma$ is very massive.  We therefore get an
effective computation on the $s^5$ plane which turns out to be very
similar to the computation done earlier on the $\sigma$ plane for
the $\bar{\bf 27}^3$ case.

We consider a family of models, parametrized by a complex parameter
$\epsilon$, such that at $\epsilon=0$,
 the potential vanishes for
$s^i=(0,0,0,0,s^5)$ and $p=cs^5$.   For small, non-zero $\epsilon$,
 $s^1,\dots,s^4$ will
still be hugely massive in the regime of large $s^5$.
For $s^i=(0,0,0,0,s^5)$ and $p=cs^5$, the only relevant
contributions to the $J_i$ are those proportional to $(s^5)^5$;
moreover, these vanish if $\epsilon=0$ and in general are proportional
to $\epsilon$.  So in this region we have $J_i=p\tilde J_i=\epsilon K_i(s^5)^5$
for some constants $K_i$,
and $J_0=\epsilon K_0(s^5)^5$ for some constant $K_0$.
With these simplifications near the singular locus,
we will find that the computation
of the ${\bf 27^3}$ coupling in this regime becomes isomorphic to that of
the ${\bf\bar{27}^3}$ coupling evaluated near t=0 in section 3.
Recall that we reduced that computation to a Hamiltonian computation
involving wavefunctions $\Psi$ annihilated by $\bq$ and $Q_+$.
In the regime of interest here (large $s^5$ and $p$) the $\s$ and
gauge multiplets are very heavy and can be set to zero.  Similarly
we only have one bosonic zero mode $s^5$ and the $s^1,\dots,s^4$
multiplets are massive.  This leaves the following expressions for
$\bq$ and $Q_+$:

\eqn\Qpsi{\eqalign{
&\bq =-\sqrt{2}\bar\psi^I_{+}{\partial\over{\partial\bar \phi^I}}
+J_I\psi^I_{-}
\equiv -\sqrt{2}\bar\alpha_+{\partial\over{\partial\bar s^5}}
+\epsilon (s^5)^5\alpha_-
\cr
&Q_+=\sqrt{2}\psi^I_{+}{\partial\over{\partial \phi^I}}
+\bar J_I\bar\psi^I_-
\equiv\sqrt{2}\alpha_+{\partial\over{\partial s^5}}
+\bar\epsilon (\bar{s}^5)^5\bar\alpha_-}}

\noindent where $\alpha_+=\psi^0_{+}/c+\psi^5_+$ and
$\alpha_-=K_I\psi^I_{-}$ (and similarly for the complex
conjugates).  This only depends on $s^5$ as it should.

To make the isomorphism with the ${\bf \bar{27}^3}$ computation
manifest, we change variables from $s^5$ to
$U(s^5)=(s^5)^6/6$
and from
$\alpha_{\pm}$ to $\chi_-=-\alpha_-{\partial U\over{\partial s^5}}
=-\alpha_- (s^5)^5$ and
$i\chi_+={\partial U\over{\partial s^5}}\alpha_+=(s^5)^5\alpha_+$.
Then

\eqn\Qchi{\eqalign{
&\bq=\sqrt{2}\bar\chi_+i{\partial\over{\partial\bar U}}
-\epsilon\chi_-
\cr
&Q_+=\sqrt{2}\chi_+i{\partial\over{\partial U}}
-\bar\epsilon\bar\chi_-}}

\noindent  These equations are isomorphic to \Qqm.  We
can now look for a wavefuntion
$\Psi(s^5, \bar s^5, \bar\chi_-, \bar\chi_+)$
which is annihilated by $\bq$ and $Q_+$.  So
correlation functions of three ${\bf 27}$'s (quintic
polynomials) with a term proportional to $(s^5)^5$ will have a simple

pole at the conifold singularity, a result derived on the (2,2) locus
by Candelas at large radius \candelas, and by Vafa at LG \toplg.

Now let us consider our real interest:
the ${\bf S^3}$ couplings.  Once again the simplest direct
approach is to compute in the half-twisted model with spins shifted
as in table 2.  In particular $p$ is twisted
(and even if one adds a multiple of $Q$ in the twisting, which could occur
dynamically if the vertex operators are enveloped by a fractional
instanton, $p$ or $s^5
$ is twisted).
Vertex operators for the singlets
(killed by $\bq$) were given in \Rvert~ and \Svert.

%\eqn\singlvert{V_B=5pH_4^a(s_i)\psi_{-,a}+S_5(s_i)\psi_{-,0}}

%\noindent which have left and right U(1) charges (0,1),
%where $H_4$ and $S_5$ are quartic and quintic polynomials,
%respectively, satisfying $s^i\delta_{ia}H_4^a=S_5$.
They have U(1) charges (0,1) and the ${\bf S^3}$ coupling
is given by a correlation function of three such vertex operators
with no extra insertions of spectral flow generators.
Again there can be no divergence, this time because
of the twisting of $p$.  We thus confirm at this singular locus what we learned
at the other: the superpotential is flat.

\subsec{Large Radius Analysis: Worldsheet Instantons}

The rest of this paper has a somewhat different emphasis.
We  want to show that under rather certain assumptions,
any pole in the superpotential of a $(0,2)$ model can only
arise by setting the complex structure or Kahler 
moduli to particular values; one  
cannot get a pole that arises upon adjusting the bundle moduli
on a generic smooth Calabi-Yau with generic metric.   
We hope that in future this will be useful in understanding
the behavior of $(0,2)$ models. 

The discussion will be carried out by looking at the sum over
worldsheet instantons.  If the superpotential has a singularity
for generic values of Kahler and complex structure
moduli (but some special bundle
moduli) this singularity must come not from summing over the worldsheet
instanton number; rather, a singularity must arise in the worldsheet instanton
contributions for some specific values of the worldsheet instanton number.
We will explore the conditions under which this might happen.
These will be practically
the only remarks in the present paper that are not limited to those
models that are associated with linear sigma models.

In the large radius limit, $(0,2)$ models are defined by choosing
(on some Calabi-Yau manifold $X$) a solution of the Kahler-Yang-Mills equations
\eqn\pino{g^{i\bar j}F_{i\bar j}=0=F_{ij}=F_{\bar i\bar j}.}
Here $g$ is the Calabi-Yau metric and $F$ is the Yang-Mills field strength.
We would like to make some simple remarks about how solutions of this
equation can develop singularities.  These remarks are not limited to the
case that $X$ has $c_1=0$, though that is where we will apply them.

First of all, we consider the case that the complex dimension of $X$ is one.
Then \pino\ says that the gauge field is flat; in particular, it can be gauged
away locally and cannot develop any singularities at all.  This simple
statement for ${\rm dim}_{\bf C}(X)=1$ is a special case of a more general
statement: a family of solutions of \pino\ cannot develop a singularity
in complex codimension one.

Now we move on to the case that the complex dimension of $X$ is two.
Then \pino\ is equivalent to the Yang-Mills instanton equations.
In this case, it is very familiar that even for $X= {\bf C}^2\cong {\bf R}^4$,
a singularity can develop as an instanton shrinks to zero size.
The singularity arises at an isolated point,
which has complex codimension two.  To achieve this singularity,
in the usual rotation-invariant description of instantons on ${\bf R}^4$,
one parameter, the instanton scale size, is adjusted.  If a complex structure
is chosen to identify ${\bf R}^4 $ as ${\bf C}^2$, the scale size
becomes the absolute value of a complex variable.  Thus, by adjusting
one complex parameter in the parameter space of solutions of \pino,
one can produce a singularity of the solution
that arises in complex codimension two on $X$.

More generally, for $X$ of any complex dimension, in a one-parameter
family of solutions of \pino, one will generically meet singularities
that will arise on $X$ on any complex codimension $\geq 2$.  In the last
paragraph, we noted a familiar example of a singularity in complex
codimension two; the quintic, as we saw in section 5.1, gives a simple
example where a singularity develops in complex codimension three.

Now let us consider a -- fallacious -- attempt to prove that in any
$(0,2)$ models, the contribution of a given worldsheet instanton
 can never develop a singularity as the 
bundle parameters are varied for generic values of the complex structure
and Kahler moduli of $X$.

Call the instanton $C$.  It is a holomorphic curve in $X$ and is
of genus zero.  We will suppose that $C$ is isolated.  (For example,
for generic complex structure on the quintic, the worldsheet instantons
are all isolated.)
For the contribution of $C$ to develop a pole as the gauge bundle
is varied, keeping the complex structure of $X$ fixed,
the gauge bundle must develop a singularity on $C$.
Otherwise, the evaluation of the contribution of the given instanton
is manifestly finite.

Now, let $n$ be the complex dimension of $X$.  $C$ has complex dimension
1 or codimension $n-1$. From the general discussion of solutions of \pino,
as a parameter is varied singularities will develop on a submanifold
$Y$ of codimension $\geq 2$.  Since $(n-1)+2>n$, it would appear that
generically we should expect the intersection of $C$ and  $Y$ to be empty.
Then the contribution of $C$ cannot generate a pole as as it does
not ``see'' the singularity.

What is wrong with this argument?
The only
fallacy  is that $Y$ may not be sufficiently generic to
allow such simple dimension counting.

Let us spell this out more precisely when $X$ has complex dimension three.
Then $Y$ has dimension one or zero; consider first the case that the dimension
is one.
With $C$ and $Y$ both of dimension one, and $1+1<3$, we would
not expect $C$ and $Y$ to meet.  In fact, dimension-counting suggests
(though it is perhaps hard to prove) that on most Calabi-Yau manifolds
it is possible
to pick a complex structure such that distinct curves $C$ and $Y$ never meet.
If this is so, how can the contribution of $C$ to the superpotential
get a pole when the bundle degenerates on $Y$?
The answer -- pointed out to us by D. Morrison -- is that such
a pole can arise precisely when $Y=C$!  Thus, we get a partial
criterion for failure of conformal invariance of $(0,2)$ models:
the space-time superpotential should be expected to develop a pole,
and thus conformal invariance should be expected to fail, if the
gauge bundle can degenerate on a curve of genus zero.

While curves on Calabi-Yau three-folds are generically isolated,
points are always free to move.  With this in mind, consider the other
possibility, that $Y$ is a point, of dimension zero.  In this case,
consider a generic one-parameter family of gauge bundles with singularities
developing on isolated points.  If all is sufficiently ``generic,''
one would expect that in a generic degeneration, $Y$ would not be located
on $C$ and therefore  the contribution to the superpotential from $C$ could
not diverge.  However, complex geometry sometimes plays tricks, and for
all we know there may be Calabi-Yau manifolds and gauge bundles such
that the singularities that arise in one-parameter families always
land on rational curves.

Anyway, we get our criterion for conformal invariance of a $(0,2)$
model defined on a Calabi-Yau threefold: if the singularities in a generic
one-parameter family of gauge bundles lie either on curves of genus
$\geq 1$ that do not meet curves of genus zero, or on points that  generically
do not lie on curves of genus zero, then   the space-time superpotential
cannot have a pole for generic values of the Kahler and complex structure
moduli.  (At special Kahler moduli, there may be a pole from summing
over instantons, and at special complex structure moduli, a pole may
come if $C$ is contained in, or at least intersects, singularities of $X$.)

This criterion is easy to implement for the quintic.  As we saw in section
5.1, in a generic family, the gauge bundle degenerates only on isolated
points, which moreover can move freely.  So our criterion is obeyed.
 It seems very likely that the same reasoning will work
for a large class of $(0,2)$ models derived from linear sigma models,
but we will not analyze this here.

\bigskip
\noindent{\it More Detail For The Quintic}

Let us describe this in more detail.  Consider a generic
one parameter family of quintic models leading (as in section 5.1)
to an isolated singularity which we may as well take to lie at
$x=(0,0,0,0,1)$.  This means that the five functions $\tilde J_i$
all vanish at $x$.  We want to show that for generic such $\tilde J_i$,
there is no rational curve on $X$ that passes through $x$.
In general, a rational curve is a map from ${\bf CP}^1$, with homogeneous
coordinates $(u,v)$, to $X$; the homogeneous coordinates $s^i$ of $X$
are homogeneous functions of $u,v$ of some degree $k$, called the degree
of the instanton.  We will for simplicity  take $k=1$, but the counting
works similarly in general.  There is no essential loss in assuming
that a curve that passes through $x$ does so at $u=0$, and then
the curve takes the form
\eqn\husn{s^i(u,v)=(\alpha_1 u,\alpha_2 u,\alpha_3 u,\alpha_4u,v).}
(We imposed that $s^i(0,v)=(0,0,0,0,v)$, and added to $v$ a multiple of
$u$ to ensure $s^5=v$. The $\alpha_i$ cannot be all zero or \husn\ would not
define a curve at all.)
\husn\ gives a linear map from ${\bf CP}^1$ to ${\bf CP}^4$ that
meets $x$; we must determine if it lies on $X$.  Setting $G= s^i\tilde J_i$
(so that $X$ is defined by $G=0$), the curve lies on $X$ if and only if
\eqn\gusn{G(s^i(u,v))=0.}
Note that $G(s^i(u,v))$ is a homogeneous function of $u,v$ of degree five.
The condition that $\tilde J_i=0 $ at $x$ means that $G(s^i(0,v))=0$,
but $G$ is otherwise generic.    So we can expand
\eqn\busso{G(s^i(u,v))=u^5P_5(\alpha_i)+u^4vP_4(\alpha_i)+\dots +
uv^4P_1(\alpha_i).}
For $G(s^i(u,v))$ to be identically zero is five equations for the
four $\alpha_i$, so generically there is no solution.  The essential
point is that the condition that the curve passes through $x$ eliminates
two parameters that could otherwise be added to \husn\ (plus more that
can be rotated away by linear transformations of $u,v$), but the condition
that $\tilde J_i$ are all zero at $x$
eliminates only one term from \busso, namely the coefficient of $v^5$.
So in asking that there is a singularity and that the curve passes through
it, we have eliminated two variables and only one equation.
Without these conditions, the curves on the quintic are generically isolated;
with the conditions, they generically do not exist.

\subsec{Localization of the Half-Twisted Path Integral}

In the above subsection, we showed that generically, the instanton
does not meet the possible solutions of the equations $\tilde J_i=0$ and
thus the possible flat directions of the potential.
Actually, one can also argue directly that nothing goes wrong
to instanton computations in the linear sigma model
even if an instanton  does meet a solution of
$\tilde J_i=0$; in fact, in the instanton sector we can even set
$\tilde J_i$
to be identically zero without producing a singularity.  The reason
is that the computation of instanton contributions localizes, as explained
in \phases, section 3.4, on the space of vortex solutions of an abelian
Higgs model that also obey $J_i=0$.  For $(0,2)$ models the
fixed point locus is determined by setting $\{\bq,F\}\equiv 0$
for all fermion fields $F$.  As long as the instanton number
is non-zero, either $p$ or all the $s^i$ must vanish in the instanton solution
because of having the wrong sign of the charge.

Once that is given,
the space of vortex solutions remains compact no matter how the
$\tilde J_i$
degenerate -- even if one takes $\tilde J_i$ to zero.
(This is somewhat analogous to the fact that in the $(2,2)$ case,
instanton computations can be performed at $G=0$ \morrison.)
Hence the instanton calculation can never develop a pole for
given instanton number.

Of course, the argument just given uses the structure of the
linear sigma model while the earlier argument that singularities
require that the gauge bundle should degenerate in complex codimension
two is valid for arbitrary $(0,2)$ models constructed from a Calabi-Yau
manifold with a holomorphic vector bundle.

%\eqn\LGinst{\eqalign{
%&s_i=0
%\cr
%&(D_1+iD_2)p=0
%\cr
%&v_{12}=e^2(-5|p|^2-r)}}

%\noindent while the singularity is produced by configurations with
%p and some $s_i$ large.  The first equation in \LGinst~ shows that
%these two regimes do not overlap.
\newsec{Discussion}

%--make sure it applies to whole class of LSMs, so we can make precise
%statement of conclusion

Although we concentrated on the quintic vacuum, it is clear that our
arguments apply more generally.  Any linear sigma model describing
the parameter space of a complete intersection in a
toric variety
will have bosonic fields charged under the $U(1)$ $R$-symmetry in such
a way that the
argument of section 4.3 will go through.  Similarly, at a generic
point on the singular locus the vector bundle $V$ will be singular
but the manifold will be smooth.

This result is surprising from the point of view of worldsheet instantons:
it can be taken as a prediction that the contributions to singlet
correlations of the $2875$ rational
curves on the quintic cancel.
At the Landau-Ginzburg end, our conclusion also implies that
{\it all} $E_6$ singlets that correspond
to deformation-theoretic modes at large radius
must have flat superpotential at small radius; this includes amplitudes
which have no symmetry reason to vanish \dksecond.
In \mirgep, this has been verified for one such amplitude
involving singlets at $r=-\infty$ which appear to correspond
to $24$ of the $224$ $H^1(End~V)$ modes at large radius.

The next natural question is to consider giving gauge {\it non}-singlet
scalars vacuum expectation values, breaking the gauge group
down to $SO(10)$ or $SU(5)$.  Obtaining, for example, an
$SO(10)$ theory by turning on the ${\bf 1}_2$ component
of the ${\bf\bar{27}}$ and the ${\bf 1}_{-2}$ component of the ${\bf 27}$
does not simply involve adding polynomial deformations to the
linear sigma model.  However, Distler and Kachru have
introduced linear sigma models
describing $(0,2)$ models which are {\it not} deformations
of $(2,2)$ models \dkfirst.
These have smaller gauge groups at the string scale.  For these
cases one must first check that the nonlinear sigma models to
which they reduce at large radius are solutions of the low-energy
equations of motion.  This ensures that the superpotential does
not diverge at infinite radius.

Then the methods used in this
paper for the quintic can be invoked to argue for the conformal invariance of
the infrared limit of
these more realistic models.
The twisting of the fields by their $U(1)_R$ charges
ensures that the singularities of the linear sigma model do
not lead to poles in singlet correlation functions.
Similarly,
the worldsheet instanton computations will not ``know'' about the
singular locus of the vector bundle.
%What about more general models, models without geometrical phases,
%etc....?
It should now be clear how to apply our considerations to still more general
linear sigma models (such as models with no large radius phase at
all \rigid):  (i) check that the fields that could potentially go to infinity
when the model becomes singular have nonzero $U(1)_R$ charges,
or (ii) study the instanton
expansion about a known locus on the parameter space and check
whether the instanton moduli space is compact and thus avoids the
singularity.

In addition to fixing the space-time superpotential for the
singlets, the methods used here might help compute other
quantities of interest for $(0,2)$ vacua.
In principle one can reliably compute quantities
which are renormalization-group
invariant and holomorphic up to surface terms
in the linear sigma model.  In practice this is difficult except
in certain limits where the computation becomes semiclassical.
As we have seen, one such limit is near the singularities.
The orders of the
poles in the Yukawa couplings of gauge-charged modes near
the singularities can be
computed for more models.  Another quantity of great interest is
the one-loop gauge coupling function, which bears on the unification
scale in string theory and the moduli-dependence of non-perturbative
effects.  Determining its behavior near the singularities using
the linear sigma model might go a long way toward fixing its dependence
on the moduli (including $(0,2)$ moduli).

\vskip .5in
\centerline{{\bf Acknowledgements}}
We thank J. Distler, S. Kachru, R. Plesser, and N. Seiberg for discussions.
E.S. is supported by an N.S.F. Graduate Fellowship and an AT$\&$T
GRPW Grant.  E.W. is supported in part by N.S.F. Grant
PHY92-45317.

\listrefs
\end